# Exciton-Coupled Coherent Magnons in a 2D Semiconductor

Youn Jue Bae[1], Jue Wang[1], Allen Scheie[2], Junwen Xu[3], Daniel G. Chica[1], Geoffrey M. Diederich[4,5], John Cenker[4], Michael E. Ziebel[1], Yusong Bai[1], Haowen Ren[3], Cory R. Dean[6], Milan Delor[1], Xiaodong Xu[4], Xavier Roy[1], Andrew D. Kent[3], Xiaoyang Zhu[1,*]

[1] Department of Chemistry, Columbia University, New York, NY 10027, USA
[2] Neutron Scattering Division, Oak Ridge National Laboratory, Oak Ridge, TN 37831, USA
[3] Department of Physics, New York University, New York, NY, USA 10003, USA
[4] Department of Physics, University of Washington, Seattle, WA 98195, USA
[5] Intelligence Community Postdoctoral Research Fellowship Program, University of Washington, Seattle, WA 98195, USA
[6] Department of Physics and Astronomy, Columbia University, New York, NY 10027, USA

**ABSTRACT. Two-dimensional (2D) magnetic semiconductors feature both tightly-bound excitons with large oscillator strength and potentially long-lived coherent magnons due to the presence of bandgap and spatial confinement. While magnons and excitons are energetically mismatched by orders of magnitude, their coupling can lead to efficient optical access to spin information. Here we report strong magnon-exciton coupling in the 2D van der Waals (vdW) antiferromagnetic (AFM) semiconductor CrSBr. Coherent magnons launched by above-gap excitation modulate the interlayer hybridization, which leads to dynamic modulation of excitonic energies. Time-resolved exciton sensing reveals magnons that can coherently travel beyond 7 μm, with coherence time above 5 ns. We observe this exciton-coupled coherent magnons in both even and odd number of layers, with and without compensated magnetization, down to the bilayer limit. Given the versatility of vdW heterostructures, these coherent 2D magnons may be basis for optically accessible magnonics and quantum interconnects.**

The recent discoveries of 2D magnets [1–6] have expanded the horizon of 2D phenomena and applications. Like other 2D materials, these magnets can be stacked into van der Waals (vdW) structures, leading to new, enhanced, and controllable properties [7–11]. One exciting application of 2D magnets is to exploit their spin waves (i.e. coherent magnons) [12] as energy efficient information carriers in spintronics and magnonics [13,14] or as interconnects in hybrid quantum systems [15–17]. For spintronics and magnonics, antiferromagnetic (AFM) materials can be advantageous because the zero net magnetization makes devices insensitive to stray fields or cross-talk[18–21]. However, the absence of magnetization also makes detecting the coherent magnon particularly challenging. For hybrid quantum systems, coherent magnons may couple to other quantum modules for transduction, coherent control, and remote entanglement. Such coupling requires a net magnetization and has been realized only for FM [15–17,22,23], not the more prevalent AFM materials.

---

[*] To whom correspondence should be addressed. E-mail: xyzhu@columbia.edu

A particular opportunity arises when a 2D material possesses both magnetic and semiconducting properties, as reported recently for CrSBr [24–26] and $NiPS_3$ [27–29] that feature excitonic transitions and are A-type antiferromagnets with intralayer FM order and interlayer AFM coupling. We choose CrSBr here because of its excellent semiconducting properties,[24,25] and more importantly, the discovery of strong coupling of Wannier excitons to interlayer magnetic order [26]. Atomically-thin flakes of CrSBr can be produced via mechanical exfoliation and the bulk magnetic structure is maintained down to the FM monolayer with a Curie temperature $T_C$ = 146 K and to the AFM bilayer with a Néel temperature $T_N$ = 140 K [25]; the latter is higher than the bulk $T_N$ of 132 K. CrSBr is also a direct-gap semiconductor down to the monolayer, with an electronic gap of 1.5 eV and an excitonic gap of 1.34 eV [24]. Towards the 2D limit, the material can be with or without net magnetization for odd or even number of layers, respectively [25]. The coexistence of both magnetic and semiconducting properties implies that a spin wave may coherently modulate the electronic structure which, in a 2D semiconductor, is reflected in the dominant excitonic transitions [30,31]. Such 2D magnon-exciton coupling allows the launch and detection of spin waves from strong absorption, emission, or reflection of light in the energy range corresponding to excitonic transitions

. This is a major advantage over conventional methods to optically access magnons via i) resonant excitation in microwave spectroscopy based on microwave antenna and waveguides [13]; ii) magneto optical effects based on precise detection of light polarization rotation [12,19], or iii) symmetry changes detected in nonlinear optical spectroscopy with high power pulsed lasers [19,25].

The coupling of Wannier excitons to interlayer magnetic order in CrSBr comes from the spin-dependent inter-layer electron exchange interaction.[26] Using first order perturbation theory, we can approximate the shift in the exciton energy ($\Delta E_{ex}$) due to changes in the interlayer electron exchange interaction as $\Delta E_{ex} \propto \cos(\theta/2)^2$ where $\theta$ is the angle between the magnetic moments (M) in neighboring layers[26]. In the AFM state ($\theta = \pi$), the interlayer hybridization is spin-forbidden ($\Delta E_{ex} = 0$); in the FM state ($\theta = 0$), interlayer electron exchange interaction is the greatest and $\Delta E_{ex}$ is -20 meV [26]. The dependence of $\Delta E_{ex}$ on $\theta$ is the basis for exciton sensing of coherent spin waves. To probe the dynamical change in $\Delta E_{ex}$, we excite CrSBr with a femtosecond laser pulse with above-gap photon energy ($h\nu_1$), and probe the resulting spin waves with a femtosecond broadband pulse at a controlled distance ($d$) between the pump and probe spots (Fig. 1a and Fig. S1). We identify the magnon modes from coherent oscillations in $E_{ex}$, and corroborate this assignment with frequency-domain magnetic resonance spectroscopy. The long-lived coherent magnons and their strong coupling to excitons allow us to directly image coherent magnons propagating in the 2D surface plane.

We prepare CrSBr flakes with thicknesses ranging from bilayer to thin-bulk onto a Si wafer with a 90 nm thick $SiO_2$ layer [25,26]. In order to enhance excitonic sensing of the spin waves, we apply an external magnetic field ($B_0$) along the $c$-axis to tilt the spins. In the AFM phase, a change to $E_{ex}$ due to interlayer hybridization is $\Delta E_{ex} \propto \cos(\theta/2)^2$, with $\theta = (\pi - \alpha_0) \pm \alpha$; $\alpha_0$ is the reduction of $\theta$ caused by $B_0$ and $\alpha$ is the time-dependent modulation of $\theta$ by the coherent spin wave. Because $dE_{ex}/d\theta \propto \sin(\alpha_0 + \alpha)$, having a non-zero $\alpha_0$ enhances the oscillatory signal, which we measure as oscillation in $E_{ex}$. We excite CrSBr by $h\nu_1$ = 1.7 eV and measure the change in reflectance at a variable time delay ($\Delta t$) by a broadband probe ($h\nu_2$ = 1.3 – 1.4 eV) to obtain transient reflectance, $\Delta R/R$, where $\Delta R$ is the differential reflectance with and without the pump and R is reflectance without the pump. This energy region probes the excitonic transitions [26], as shown by a static reflectance spectrum in Fig. 1b. Each $\Delta R/R$ spectrum with above-gap photoexcitation features coherent oscillations on top of an incoherent background (Fig. S2). We isolate the pure oscillatory response by subtracting the incoherent signal attributed to electronic excitations for a thin bulk CrSBr at $T$ = 5 K and $d$ = 0 μm (Fig. 1c). We choose the pump laser power in the linear electronic excitation region with sufficient oscillatory amplitude (Fig. S3).

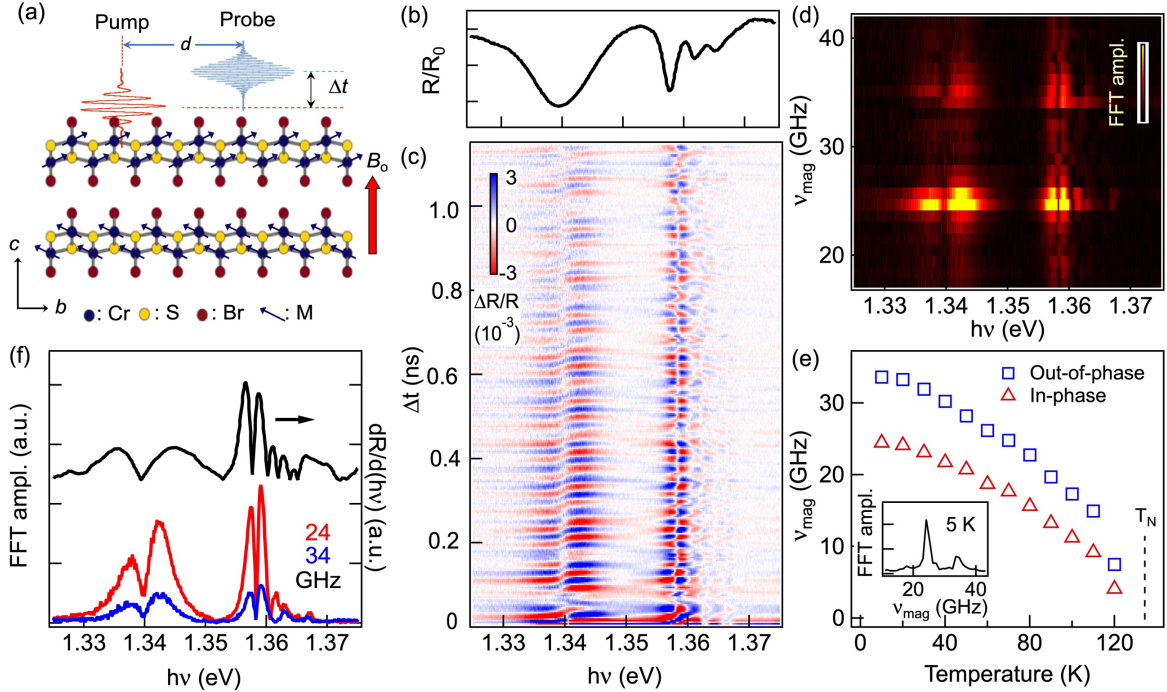

**Fig. 1. Coupling of excitons to coherent magnons.** (a) Detecting spin waves from transient reflectance. A femtosecond pump pulse with above gap photon energy launches spin waves in the layered AFM semiconductor CrSBr. Reflectance from a broadband probe pulse measures coherent oscillation at excitonic transitions. An external canting field, $B_0 = 0.2$ T, is applied along the c-axis (normal to the 2D planes), to enhance the oscillatory signal. (b) Reflectance ($R$) from CrSBr is normalized to that of the SiO$_2$ substrate ($R_0$). (c) Transient reflectance spectra $\Delta R/R$ as a function of pump-probe delay ($\Delta t$) and probe photon energy ($h\nu$). An incoherent background has been subtracted from $\Delta R/R$ (see Fig. S2). The pseudo color scale is $\Delta R/R$, where R is reflectance without pump and $\Delta R$ is pump-induced change in reflectance. (d) 2D fast Fourier transform (FFT) of the data in panel (c). The pseudo color (normalized, 0-1) is the FFT amplitude. (e) The two spin wave frequencies are shown as a function of sample temperature below $T_N$. The inset is the probe $h\nu$-integrated FFT trace showing the two peaks at 24 and 34 GHz at 5 K. (f) Two horizontal cuts of the 2D pseudo-color plot in panel (d) at the two peak frequencies (red, 24 GHz; blue 34 GHz). Also shown is the first derivative (black) of the reflectance spectrum in (b). All experiments are done at a sample temperature of 5 K, except panel (e) where the sample temperature is varied between 5 and 140 K. The pump pulse at $h\nu_1 = 1.7$ eV (pulse width: 150 fs, rep-rate 250 kHz, power 1 μW, spot diameter: 1.4 μm) and the probe pulse at $h\nu_2 = 1.3 – 1.4$ eV (pulse width ~100 fs, power 0.2 μW, spot diameter ~1.2 μm) are used.

The oscillatory components must come from the coherent spin waves because (i) they are observed only below $T_N$ and the frequencies decrease with $T$ (Fig. S4 and Fig. 1e discussed below); and (ii) the amplitude of oscillation signal is much weaker in the absence of a canting field (Fig. S5) and increases with $B_0$. To obtain the frequencies of the coherent spin waves, we perform a fast Fourier transform (FFT) of the oscillatory response. Fig. 1d plots the FFT amplitude (pseudo color) as a function of spin wave frequency ($\nu_{mag}$) and $h\nu_2$. There are two peak frequencies at ~24 GHz and ~34 GHz, both independent of probe $h\nu_2$. This is expected because the coherent spin waves modulate the interlayer electronic hybridization, and thus all excitonic transitions. We use the $h\nu_2$-integrated FFT spectrum (inset of Fig. 1e for a typical spectrum at $T = 5$K), to quantify the peak frequencies: $\nu_{mag1} = 24.6 \pm 0.7$ GHz and $\nu_{mag2} = 34 \pm 1$ GHz. With increasing $T$, the magnetic order decreases and this results in lowering of the spin wave frequencies (Fig. 1e). Around $T_N$, both $\nu_{mag1}$ and $\nu_{mag2}$ approach zero. The T-dependences of $\nu_{mag1}$ and $\nu_{mag2}$ closely follow that of the magnetic order parameter [25].

The strong coupling of excitons to coherent magnons is revealed by clear π-phase flips of the oscillatory signal at $h\nu_2$ corresponding exactly to the peaks of excitonic transitions, (Fig. 1b, c). The π-phase shift is a signature of an optical transition modulated by coherent oscillation in a coupled mode [32], as is also known for coherent phonon-exciton coupling [33,34]. Likewise, the FFT amplitudes of both $\nu_{mag1}$ and $\nu_{mag2}$ (respectively red and blue spectra in Fig. 1f) track the first derivative (black) of the static reflectance spectrum. The strength of the coupling between excitons and coherent magnons can be calculated by the modulation of the exciton energy (detailed in SI and Figs. S6-12). We obtain exciton energy modulation caused by coherent magnons of $\delta E_{ex} = 4.0 \pm 0.5$ meV. Note that this coupling is perturbative in nature, not due to resonant hybridization between two modes that differ in energies by four-orders of magnitude: $E_{ex} \sim 1.3$ eV and $h\nu_{mag} = 0.10$ and $0.14$ meV (for $\nu_1$ and $\nu_2$, respectively).

To support assignments of coherent magnon modes from excitonic sensing, we use magnetic resonance spectroscopy. Fig. 2a shows a series of magnetic resonance spectra at selected microwave frequencies ($\nu = 5$-$21$ GHz) for bulk CrSBr at $T = 5$ K. The spectra reveal a single resonance in the low frequency ($\leq 18$ GHz) region and two resonances in the high frequency ($> 22$ GHz) region. We extract peak frequencies ($\nu_{mag}$) of the resonances (Fig. S13) and plot them as function of the magnetic field applied along the $c$-axis ($B_0$) (Fig. 2b). For $B_0$ smaller than a saturation field ($B_{sat} \sim 1.7$ T), we observe two $\nu_{mag}$ branches whose frequencies decrease with increasing $B_0$, consistent with reduction in AFM order as spins are progressively canted away from the easy $b$-axis. Frequencies of these two branches are assigned to the in-phase ($\nu_{IP}$) and out-of-phase ($\nu_{OP}$) spin precessions, similar to those observed in the 2D AFM materials CrCl$_3$ and CrI$_3$ [12,35,36]. Above $B_{sat}$, the spins are fully polarized parallel to $B_0$ and $\nu_{mag}$ increases linearly with $B_0$ (seen here for the out-of-phase magnon), which is expected for a ferromagnetic resonance [37]. In agreement with the magnetic resonance results, the $B_0$ dependence of the excitonic transitions in transient reflectance exhibits the same dispersion, Fig. 2c for the low-frequency in-phase branch. The temperature dependence of the magnon frequencies from magnetic resonance measurements (Fig. S16) are in good agreement with those obtained from optical measurements (Fig. 1f).

We quantitatively analyze the $B_0$ dependence of magnon frequencies using the Heisenberg spin Hamiltonian and fitting the magnetic resonance spectroscopy data using linear spin wave theory (LSWT) [38], with the fitting details shown in Fig. S14 (SI section 6). The triaxial anisotropy in spin exchange interaction and interlayer exchange interaction in CrSBr results in two non-degenerate magnon modes. The LSWT analysis gives the in-phase mode at frequency $v_{IP} = 2S\sqrt{A_x A_z + A_x J_{int}}$ and the out-of-phase mode at frequency $v_{OP} = 2S\sqrt{A_x A_z + A_z J_{int}}$. Here, $J_{int}$ is the interlayer exchange and $A_x$ and $A_z$ are single ion anisotropy constants. The $A_x$, $A_z$ and $J_{int}$ values from LSWT fits (solid curves in Fig. 2b) are 14 μeV, 58 μeV and 6 μeV, respectively. The fits give magnon frequencies at $B_0 = 0.2$ T (external field used in Fig. 1) of $v_{IP} = 24.4$ GHz and $v_{OP} = 33.8$ GHz, in good agreement with those measured from exciton sensing. Similar agreement with experimental results is also achieved in analysis based on the classical Landau-Lifshitz (LL) equation[37], with the spin configuration shown in Fig. S15 (SI, section 7).

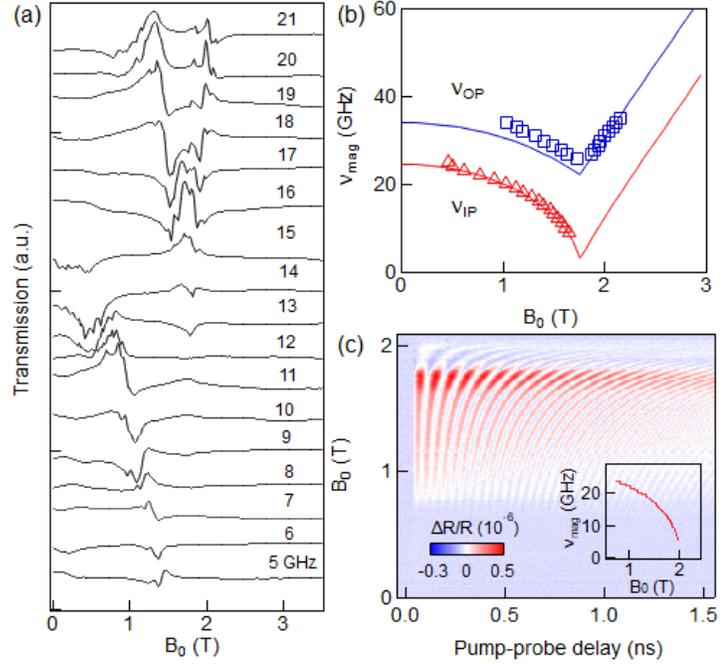

**Fig. 2. Magnetic-field dependent magnon frequencies and calculated dispersions.** (a) magnetic resonance spectra at the indicated frequencies (5-21 GHz) from a bulk CrSBr crystal at $T = 5K$, with the magnetic field applied along the $c$-axis. The spectra are offset for clarity. (b) Peak resonance frequencies as a function of magnetic field (symbols). Solid curves are dispersion fits from LSWT as detailed in SI section 7. (c) Transient reflectance spectra at a single probe wavelength with respect to the applied field along the $c$-axis of the bulk CrSBr crystal at 5 K. The inset shows the extracted peak frequency from the FFT spectrum as a function of magnetic field.

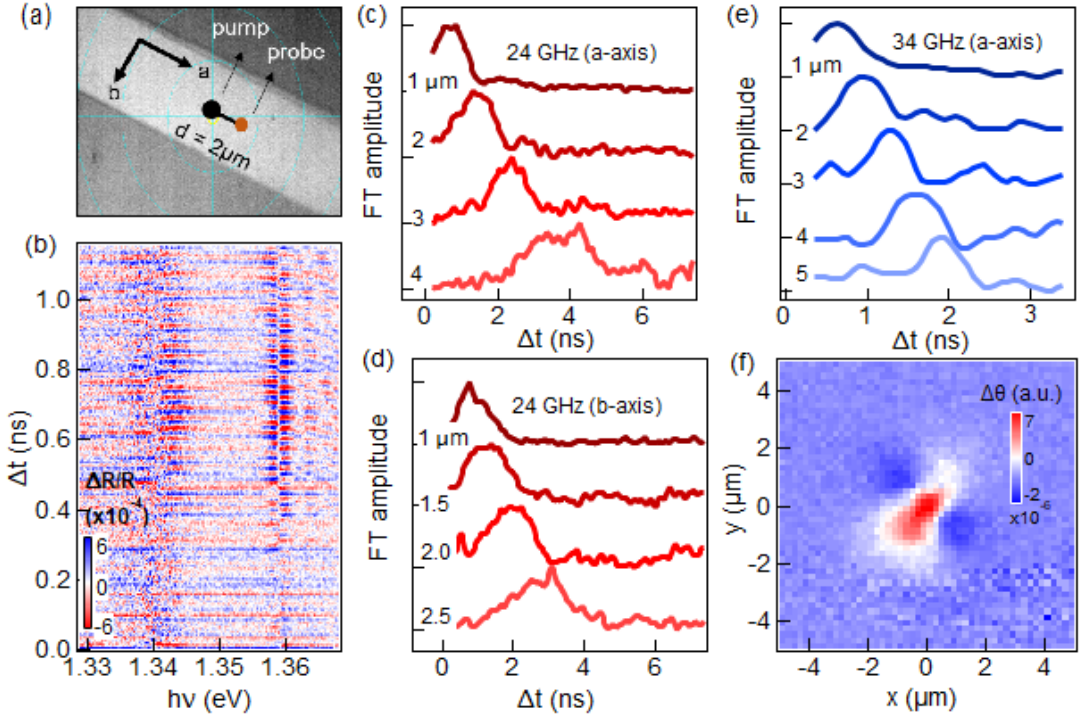

**Fig. 3. Exciton sensing of propagating coherent AFM spin waves in the 2D plane.** (a) Optical image of a CrSBr bulk crystal with pump and probe pulse spatially separated by $d = 2$ μm. (b) Transient reflectance spectra of the measurement shown in (a) ($d = 2$ μm) from which the incoherent electronic signal was subtracted. (c-f) Probe-wavelength integrated STFT spectra at different pump-probe distances for the (c) 24 GHz mode along the a- and b- axis and (d) 34 GHz mode along the a-axis. The moving time window of 0.15 ns and 0.6 ns are used for magnon and acoustic phonon modes, respectively. The same pump and probe laser conditions as in Fig. 1 are used and the sample temperature is 5 K. (g) spatially resolved MOKE imaging at pump-probe delay time of 100 ps.

We exploit the efficient exciton sensing to implement time and spatially resolved spectroscopic imaging to investigate propagating coherent spin waves. We perform this with the pump and probe beam separated by a controlled distance ($d$), Fig. 3a. The diffraction-limited excitation spot can create a gradient in the driving force for launching coherent magnons in a finite momentum window at $|k| > 0$. The propagating nature of the coherent spin wave is clearly seen in the delayed rise of the oscillatory signal for $d > 0$, as illustrated for $d = 2$ μm along the $a$-axis, Fig. 3b (see Fig. S17 for probe photon energy resolved propagation images). This delayed response is in stark contrast to the prompt rise of the spin wave signature when the pump and probe pulses overlap spatially ($d = 0$) in Fig. 1c. FFT analysis of the propagating waves reveals the two coherent spin waves at 24 and 34 GHz. Along the $a$- axis, we detect two magnon frequencies (24 and 34 GHz); along the $b$-axis, we only detect the 24 GHz component. We perform short-time FT (STFT) with respect to Δt for different $d$ values for the 24 and 34 GHz modes along the $a$-axis (Figs. 3c-d) and for the 24 GHz mode along the *b*-axis (Fig. 3e). In each case, the peak Δt shifts linearly with *d*, establishing the propagating nature of these coherent waves. From these shifts, we obtain group velocities of $V_g$ = 1.0 ± 0.1 and 3.0 ± 0.3 km/s, for the 24, 34 GHz along the *a*-axis, respectively and $V_g$ = 0.7 ± 0.1 for the 24 GHz mode along the *b*-axis. (Fig. S17-S18). From the probe distance-dependent measurements along the *b*-axis (Fig. S19), we obtain lower bounds for the coherent transport lengths of $\lambda_{coh}$ = 6 and 7 μm for the 24 and 34 GHz modes, respectively. In complementary experiments, carry out time-resolved imaging based on the magneto-optical Kerr effect (MOKE). Following initial excitation in the center of the image frame, the MOKE responses expand in space with Δt, as expected from the propagating nature of the spin waves (detailed in SI, section 2). Fig. 3f shows a representative early time (Δt = 0.1 ns) MOKE image, which will be discussed below.

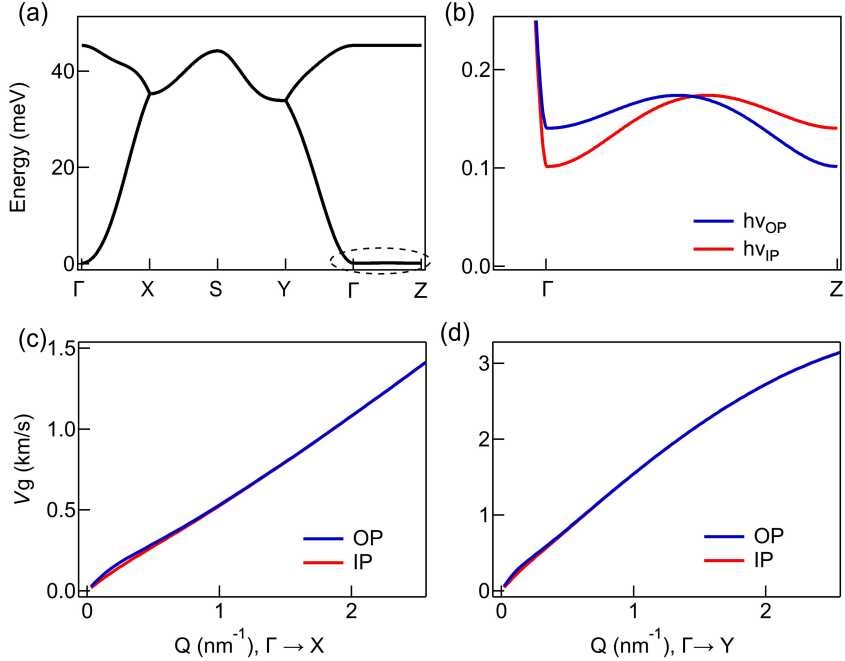

**Fig. 4. Magnon dispersions.** (a) Calculated magnon dispersions [38]. Note the dispersions from the weak interlayer spin exchange in the Γ-Z direction (dashed oval) are nearly flat. (b) A magnified view of the Γ-Z direction showing the dispersions of in the out-of-phase (OP, blue) and in-phase (IP, red) modes. (c) and (d) momentum dependence of magnon group velocity along Γ→X and Γ-Y directions, respectively.

To understand the origin of propagating spin waves, we show calculated magnon dispersions along high symmetry directions (Fig 4a, b), obtained from LSWT fits to neutron scattering data[38], taking into account the small interlayer spin exchange in the vertical direction (*c*-axis) and anisotropy exchange constants from fitting the magnetic resonance spectroscopy data in Fig 2b.

The dispersions along the in-plane Γ-X (*a*-axis) and Γ-Y (*b*-axis) directions are determined by the strong FM exchange interaction. Along the Γ-Z (*c*-axis) direction, dispersion is essentially flat on the energy scale for Γ-X and Γ-Y and becomes visible only when we zoom-in by two orders-of-magnitude, Fig. 4b. The weak interlayer AFM couplings give rise to vertical bandwidths of only ~70 μeV, as compared to ~40 meV in the in-plane directions. From the dispersions in Fig. 4a, we obtain group velocities ($V_g$) along the Γ-X and Γ-Y directions (Fig. 4c, d). There are two reasons why the propagating spin waves observed in experiments cannot be attributed to their intrinsic properties at sufficiently high momentum vectors ($Q$): i) To reach the measured $V_g$ values of 1-3 km/s, the intrinsic spin waves in the 2D plane (a- or b-axis) must possess $Q$ values > 1 nm$^{-1}$, which is three orders of magnitude larger than the moment vector ~1 μm$^{-1}$ expected from the spatial gradient from a diffraction-limited excitation spot; ii) The calculated $V_g$ values of the IP and OP modes are nearly identical in broad $Q$ ranges along both *a*- and *b*-axes, contrary to the experimental ratio of $V_{g, OP}/ V_{g, IP}$ = 3.0±0.4 along the *a*-axis and observation of only the IP mode along the *b*-axis.

A well-known mechanism for fast propagation of magnons at small momentum vector is attributed to hybridization between magnon and acoustic phonons via magnetoelastic coupling[39–41]. In this mechanism, the above gap pump pulse creates thermally induced strain in a magnetic materials and launches the hybridized magnon and phonon modes. Confirmation of this coupling mechanism comes from the short-time MOKE image (Fig. 3f) with the distinct quadrupolar shape, which is a signature of coupling of magnons to longitudinal acoustic phonons[39,42]. Further support for this mechanism comes from peak splitting in the frequency domain (Fig S20), which be directly attributed hybridization and resulting avoided crossing between the two modes. The two different group velocities presumably originate from crossing between two different acoustic phonon branches with different dispersions in momentum space.

The detection of coherent spin waves in CrSBr from simple exciton sensing allows us to extend the measurements to the 2D limit on exfoliated flakes. Note that this approach cannot be used to probe the FM monolayer because it lacks interlayer exciton coupling [26]. Similar to findings on the bulk crystal, the transient reflectance spectra from two (2L) to five (5L) layers 2L-5L feature coherent oscillations attributed to spin waves. Fig. 5a shows spectra for 5L CrSBr obtained under the same conditions as that in Fig. 1c. There is clear π phase flip in the oscillatory signal at the exciton peak of ~1.362 eV due to the strong coupling of coherent magnons to the exciton. Fig. 5b shows a line cut at $h\nu_2$ = 1.359 eV for the longest delay within our experimental limit, $\Delta t$ = 7.6 ns. The coherent oscillation clearly persists beyond the experimental time window. Lorentzian fit to its FFT, Fig. S21, gives a full-width-at-half-maximum of 0.20±0.02 GHz, corresponding to $\tau_{coh}$ = 5 ns. Similar measurements for 2L - 4 L are shown in Fig. S9 and Fig. S12. These nanosecond spin wave coherences times are similar to those in the bulk crystal, $\tau_{coh}$ = 2.5 - 5.0 ± 0.5 ns (Fig. S19). Note that the measured $\tau_{coh}$ and $\lambda_{coh}$ represent lower bounds in the intrinsic coherence times and lengths, as the measurement with finite excitation pulse width, gradient in excitation profiles, and inhomogeneity in sample environment introduces extrinsic decoherence.

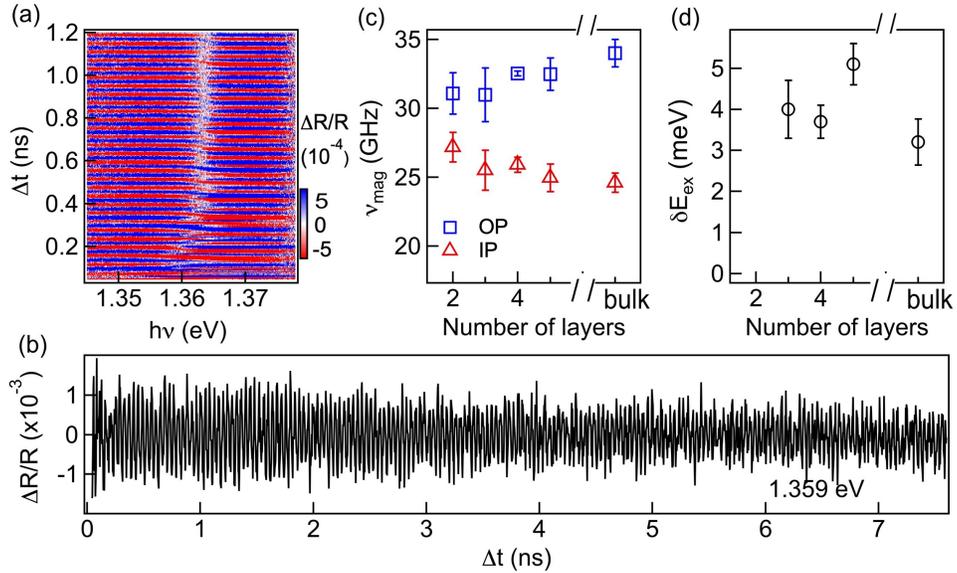

**Fig. 5. Detection of coherent spin waves from excitonic transitions in CrSBr down to the 2D limit.** (a) 2D pseudo-color plot of transient reflectance spectra from 5L CrSBr flake as a function of pump-probe delay ($\Delta t$) and probe photon energy ($h\nu_2$) at 5 K. (b) Vertical cut of the spectra in (a) at $h\nu_2$ = 1.359 eV showing long-lived coherence. (c) frequency of the in-phase (IP) and out-of-phase (OP) modes as a function of sample thickness; (d) magnitude of the exciton energy modulation with respect to the number of layers. The error bars are obtained from multiple samples and from the time average for $\delta E_{ex}$ in each sample.

Fig. 5c summarizes the dependence of the spin wave frequencies on the layer number. With increasing thickness, the frequency of the in-phase mode gradually decreases by ~10% (from 24.0 ± 1.0 to 24.6± 0.7 GHz), while that of the out-of-phase mode increases by ~10% (from 31±1 to 34±1 GHz). Fig. 5d plots the modulation in exciton energies, $\delta E_{ex}$, by the coherent magnons for layer thicknesses ranging from 3L to thin bulk. The lower signal-to-noise ratio from the 2L sample prohibits a quantitative analysis of $\delta E_{ex}$. Within experimental uncertainty, the magnon-exciton coupling energy from 3L to 5L is the same as that from the bulk, $\delta E_{ex} = 4.0 \pm 0.5$ meV (Fig. S11). An attractive attribute of the coherent magnons approaching the 2D material limit is that there can have net magnetization for odd (not even) number of layers. Having a net magnetization is essential to application as quantum interconnect [15–17,22,23]. In this regard, the sufficiently long coherence time and coherent length may meet the needs for remote coupling of qubits or quantum emitters.

Our findings of exciton-coupled coherent magnons in a 2D semiconductor represent several key advances. The coupling of coherent magnons, living in the sub-meV energy range, to excitons living in the eV region, can enable direct coupling between microwave photons and near-IR to visible photons, potentially useful for quantum transduction. This coupled exciton-magnon also offers a unique opportunity to control exciton properties of CrSBr via magnon or vice versa using optical or microwave cavities and electrostatic gating. In addition, the efficient exciton-magnon coupling opens the door to exciting applications, including integration with optoelectronics using on-chip light-emitting diodes and photodetectors. In this context, our discoveries of coherent magnons with excellent coherent properties down to the 2D bilayer limit implies that it could be combined with other 2D materials through vdW heterojunctions, leading to novel device architectures and applications.

**Acknowledgement.** The spectroscopic and imaging work at Columbia University was supported by the Materials Science and Engineering Research Center (MRSEC) through NSF grant DMR-2011738. Synthetic of the CrSBr crytals is supported as part of Programmable Quantum Materials, an Energy Frontier Research Center funded by the U.S. Department of Energy (DOE), Office of Science, Basic Energy Sciences (BES), under award DE-SC0019443. The AFMR work at NYU was supported by the Air Force Office of Scientific Research under Grant FA9550-19-1-0307. The magnetic field-dependent experiment described in Fig. 2c was performed at the University of Washington and supported by the Department of Energy, Basic Energy Sciences, Materials Sciences and Engineering Division (DE-SC0012509). The PPMS used to perform vibrating sample magnetometry measurements was purchased with financial support from the NSF through a supplement to award DMR-1751949. This research used resources at the Spallation Neutron Source, a DOE Office of Science User Facility operated by the Oak Ridge National Laboratory. G.D. acknowledges the support by an appointment to the Intelligence Community Postdoctoral

Research Fellowship Program at University of Washington, administered by Oak Ridge Institute for Science and Education through an interagency agreement between the U.S. Department of Energy and the Office of the Director of National Intelligence. We are grateful to Dr. Kihong Lee, Dr Taketo Handa and Dr Yanan Dai for providing invaluable help and insightful discussions.

SUPPORTING INFORMATION

**Exciton-Coupled Coherent Magnons in a 2D Semiconductor**

Youn Jue Bae[1], Jue Wang[1], Allen Scheie[2], Junwen Xu[3], Daniel G. Chica[1], Geoffrey M. Diederich[4,5], John Cenker[4], Michael E. Ziebel[1], Yusong Bai[1], Haowen Ren[3], Cory R. Dean[6], Milan Delor[1], Xiaodong Xu[4], Xavier Roy[1], Andrew D. Kent[3], Xiaoyang Zhu[1,†]

[1] Department of Chemistry, Columbia University, New York, NY 10027, USA
[2] Neutron Scattering Division, Oak Ridge National Laboratory, Oak Ridge, TN 37831, USA
[3] Department of Physics, New York University, New York, NY, USA 10003, USA
[4] Department of Physics, University of Washington, Seattle, WA 98195, USA
[5] Intelligence Community Postdoctoral Research Fellowship Program, University of Washington, Seattle, WA 98195, USA
[6] Department of Physics and Astronomy, Columbia University, New York, NY 10027, USA

---

[†] To whom correspondence should be addressed. E-mail: xyzhu@columbia.edu

1. CrSBr Synthesis and Sample Preparation

**Reagents**. The following reagents were used as received unless otherwise stated: chromium powder (99.94%, -200 mesh, Alfa Aesar), sulfur pieces (99.9995%, Alfa Aesar), and bromine (99.99%, Aldrich).

**Synthesis of CrBr$_3$**. Chromium (1.78 g, 34.2 mmol) and bromine (8.41 g, 105 mmol) were loaded into a bent, 19 mm O.D., 14 mm I.D. fused silica tube. The tube was evacuated to a pressure of ~30 mtorr and flame sealed while the bottom was submerged in liquid nitrogen. The end of the tube containing the chromium was heated at 1000 °C for 3 days while the other end outside the furnace was maintained below 50 °C to prevent over pressurizing the tube. *Caution: Heating the entire tube above 120°C with significant excess of bromine can result in the explosion of the tube.*

**Synthesis of CrSBr**. In a typical reaction, chromium (0.189 g, 3.63 mmol), sulfur (0.196 g, 6.11 mmol), and chromium tribromide (0.720 g, 2.47 mmol) were loaded into a 12.7 mm O.D., 10.5 mm I.D. fused silica tube. The tube was evacuated to a pressure of ~30 mtorr and flame sealed to a length of 12 cm. The tube was placed into a computer-controlled, two-zone, tube furnace. The source side was heated to 550 °C in 6 hours, allowed to soak for 12 hours, heated to 930 °C in 12 hours, allowed to soak for 84 hours, and then water quenched. The sink side was heated to 570 °C in 6 hours, allowed to soak for 12 hours, heated to 970 °C in 12 hours, allowed to soak for 24 hours, cooled to 850 °C in 12 hours, allowed to soak for 48 hours, and then water quenched. *Caution: Proper PPE should be worn when water quenching tubes including face shield, thick heat resistant gloves, and fire-resistant lab coat.* The crystals were cleaned by soaking in acetonitrile for 1 hour, followed by soaking in DI water for 16-24 hours at 70°C. After soaking, the crystals were thoroughly rinsed with DI water and acetone.

Bulk crystals were mechanically exfoliated on a Si/SiO$_2$ substrate passivated by 1-dodecanol inside a glovebox.

2. Pump-probe Spectroscopy/Microscopy Setups

Montana cryostat is used to cool down the temperature down to 4K. A permanent magnet is attached to the sample holder using epoxy glue to achieve 0.2 T at the sample position. The magnetic field at the sample position is measured at room temperature using a magnetometer (Extech SDL900). The steady state reflectance is measured using a 3200K halogen lamp (KLS EKE/AL). The broadband white light reflected from the CrSBr sample and the substrate are collected by an InGaAs photodiode array (PyLoN-IR, Princeton Instruments).

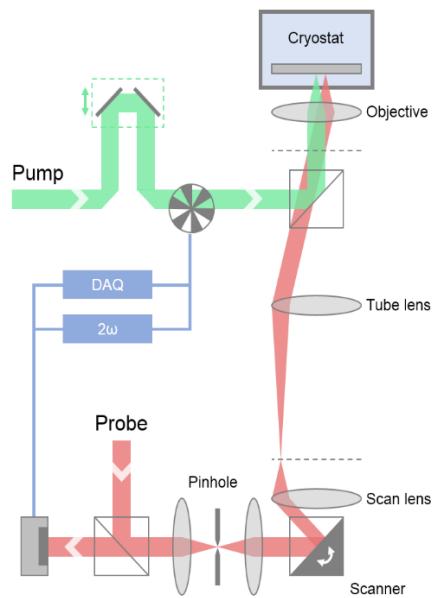

**Fig. S1.** Schematic illustration of pump-probe confocal optical microscope.

The detailed description of the transient reflectance set up is discussed previously.[1] Briefly, femtosecond laser pulses from the Ti:sapphire regenerative amplifier (Coherent RegA 9050, 250 kHz, 800 nm, 100 fs) is split into two beams: one for the tunable pump (Coherent OPA 9450) and the other for the broadband white light. A sapphire crystal is used to generate probe beam. The pump pulse is delayed using a motorized delay stage up to 7 ns and modulated using an optical copper to record pump-on and pump-off reflected signal using the InGaAs detector. The difference signal is normalized with the pump-off signal to yield ΔR/R. The pump and the probe beam get combined using a 50:50 beam splitter in front of an objective lens. A dual-axis galvo mirror scanning system is used to locate a probe pulse a few μm distance away from the pump pulse (Figure S1). The detailed description is written elsewhere.[2] Although the angle of incidence of the probe pulse at the back focal plane of an objective changes during the scanning process, the angle of incidence to the sample plane is kept normal for both the probe and pump pulse. Since the numerical aperture of the objective is high, the angle of incidence spans from -45° to 45°. Reflected pump and probe pulse travels the same path and the pump pulse is removed using a 800nm long pass filter.

**Magneto-optical Kerr effect (MOKE).** For MOKE imaging measurement shown in Fig. 3f, a pre-amplified balanced detector (PDB210A, Thorlabs), lock-in amplifier (SR830, Stanford research systems) with a chopper frequency as a reference frequency and a half-waveplate were used. A 880nm bandpass filter with the 70nm full-width-half-maximum (FWHM) was used for the broadband white light and 740 nm pump pulse from the OPA was used to excited the sample at 5K. Both the pump and probe beams pass through the same Glan-Taylor polarizer (polarized along the *b*-axis of the crystal). A 50:50 beam splitter was placed between the objective and the Glan-Taylor polarizer to direct the reflected beam into a balanced detector passing thorough a 800 nm longpass filter, half-waveplate, Wollaston prism, and a lens. Here, the probe beam travels through a dual-axis galvo scanning mirrors to precisely control the position with respect to the pump beam and to collect the pump-induced change in Kerr rotation.

**Transient reflectance with variable magnetic field.** For transient reflectance measurements, the 930 nm pulses from a 76 $MH_z$ Ti:Sapph oscillator were focused into a BBO crystal to produce 465 nm SHG which was then separated from the fundamental by a dichroic beam-splitter. The 930 probe pulses were delayed via retroreflectors on a mechanical delay line, attenuated, and polarized along the *b* axis of the sample crystal. The 465 pump pulses were sent through a mechanical chopper, attenuated, and polarized before being recombined with the probe pulses and sent to a 0.6 NA microscope objective where they were focused to a spot size of ~2 micron on the sample. The reflected pulses were picked off by an OD 3 reflective neutral density filter and sent through a Wollaston prism to separate orthogonal polarization states and each beam was passed through an absorptive long pass filter and lightly focused onto one of two detectors on a balanced photoreceiver. The signal is then sent to a lock-in amplifier, demodulating at the frequency of the mechanical chopper to ensure that only the pump-induced signal change is measured.

To generate delays and measure the dynamics, the stage is continuously swept across its range at low speed while the lock-in amplifier streamed the demodulated data to the host computer. Such scans (typically N > 25) are collected and averaged to form the final data set, with the delay calculated from the stage speed, its initial position, and the measured timebase of the lock-in amplifiers internal clock. Each photodiode on the balanced detector is sent to a different lock-in channel and their sum provides the transient reflectance.

For magnetic field measurements, the cryostat used has an integrated superconducting magnet capable of applying 5 Tesla in the *z* direction across all temperatures (base temperature is ~2K). This introduces polarization rotations due to the Faraday effect in the objective glass, which we measure and correct with half waveplates in the pump, probe, and signal beam paths.

3. Additional Spectra and Analysis of Transient Reflectance

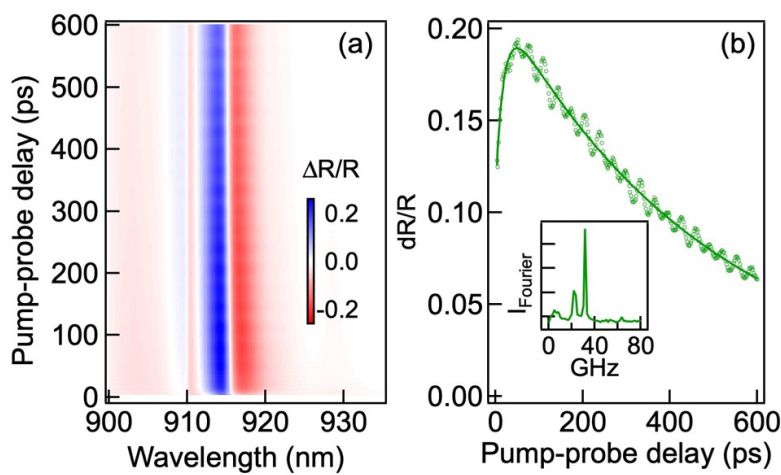

**Fig. S2.** (a) Transient reflectance spectra as a function of pump-probe delay and probe wavelength for a bulk CrSBr crystal at 5 K and with an external canting field, $B_o$ = 0.2 T applied along the *c*-axis (normal to the 2D planes. The pump pulse at $h\nu_1$ = 1.7 eV (pulse width: 150 fs, rep-rate 250 kHz, power 1 µW, spot diameter: 1.4 µm) and the probe pulse at $h\nu_2$ = 1.3 – 1.4 eV (pulse width ~100 fs, power 0.2 µW, spot diameter ~1.2 µm) are used. (b) a vertical cut of the transient reflectance in (a) at 909 nm (green circles). The solid curve is a fit of the data to a gaussian convoluted with a bi-exponential decay to account for the smooth background attributed to excitonic excitation. Subtraction of the data by the smooth fit leaves the oscillatory part, whose Fourier transform yields the frequency spectrum in the inset. The subtraction the smooth background is done at every probe wavelength to yield the $\Delta R/R$ spectra in Fig. 1c.

**Power dependence and polarization dependence.** At low power the magnon FFT amplitude linearly increases with the pump power but eventually plateaus. The pump power has a minimal effect in magnon frequency. In addition to the pump power dependence, we also use linearly polarized pump along the *b*- and *a*-axis using the same pump power (0.4 μW). Since the transition dipole is along the *b*-axis, more carriers are generated when pump beam is polarized along the *b*-axis. Thus, greater stress is also induced and we observe a greater magnon amplitude. In addition, decreased amplitude is also observed with a circularly polarized light because less carrier is generated.

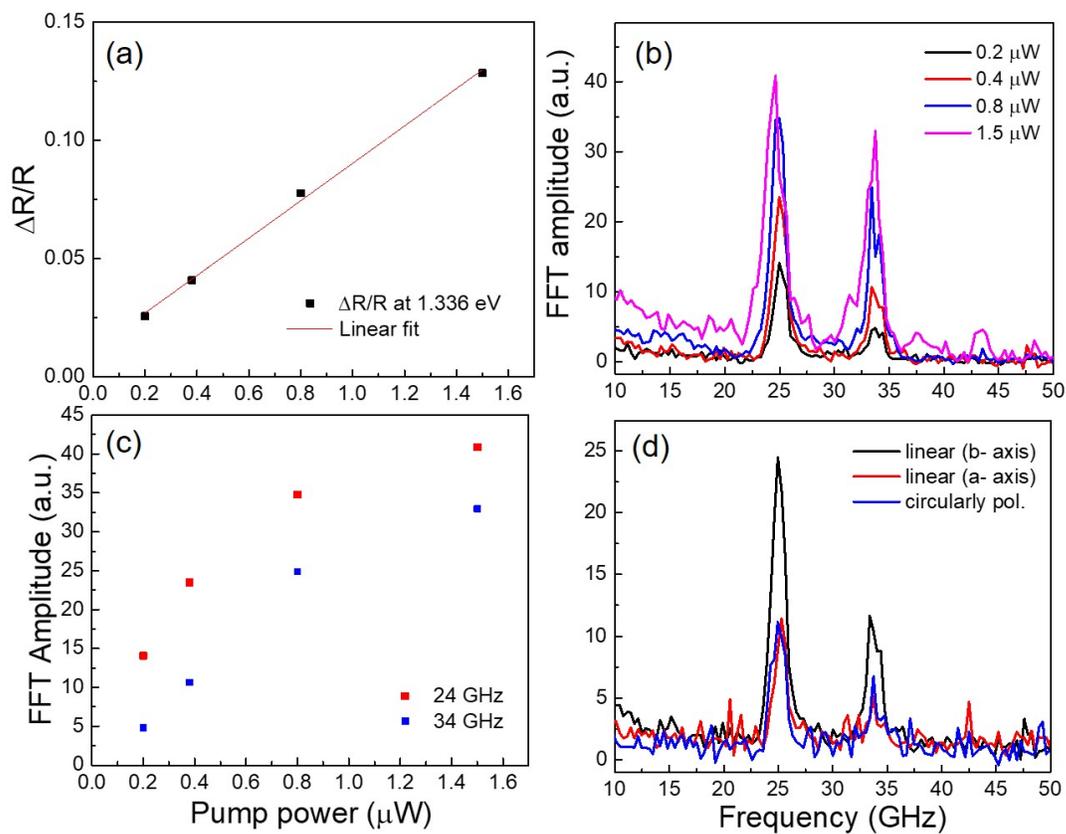

**Fig. S3**. (a) ΔR/R signal at 1.336 eV vs. pump power. (b) Linearly polarized along the *b*-axis pump power dependent FFT amplitude and (c) the maximum peak amplitude vs. pump power at 24 and 34 GHz. (d) polarization dependent FFT amplitude. Note the same pump power of 0.4 μW is used for different polarizations.

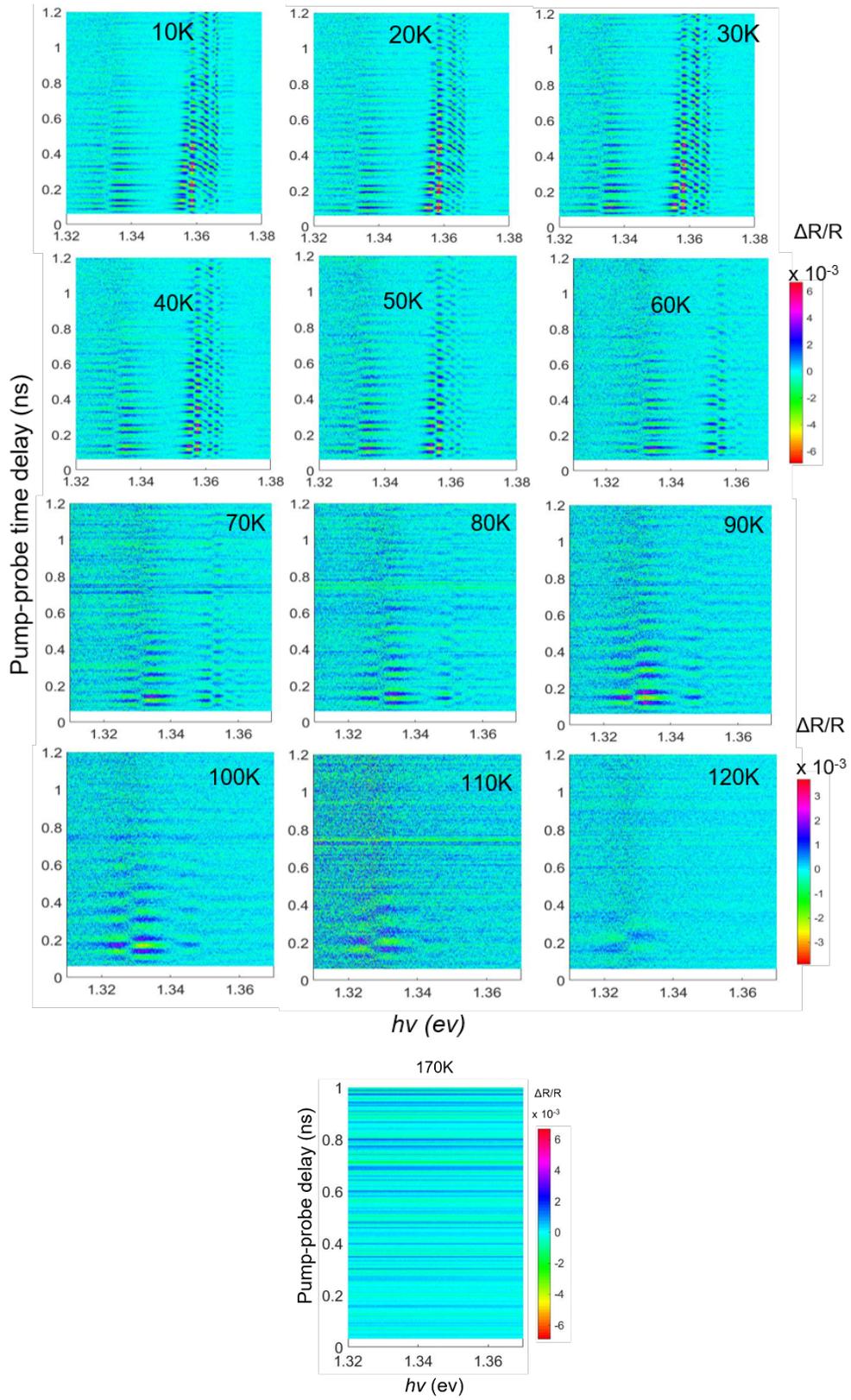

**Fig. S4.** ΔR/R spectra at different temperatures below and above the Neel temperature ($T_N$ = 135K).

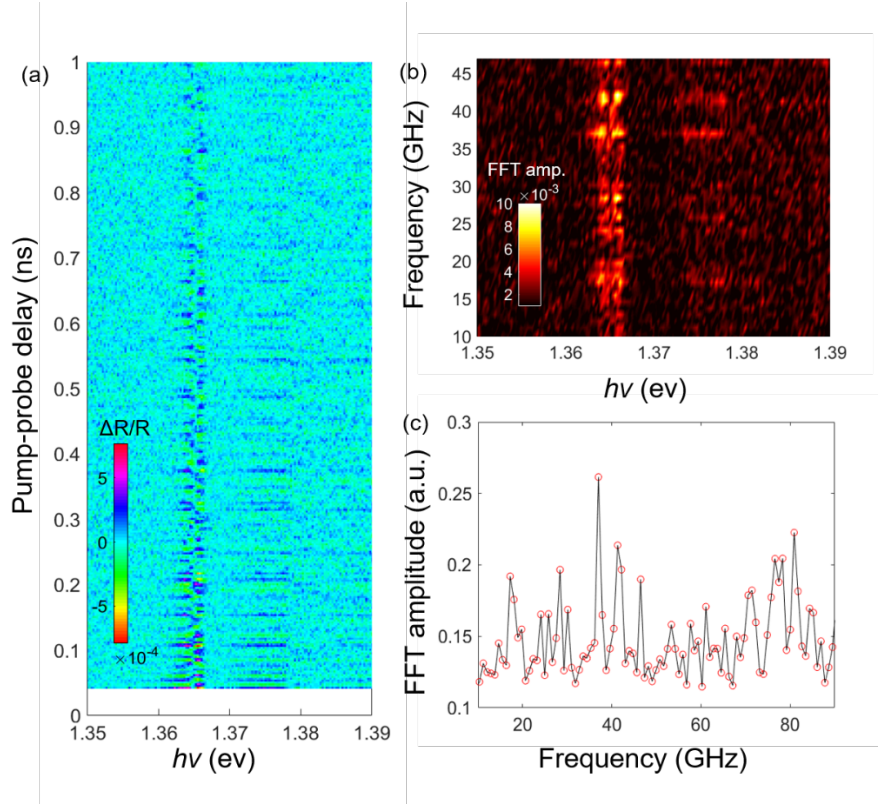

**Fig. S5.** (a) ΔR/R spectra without the incoherent electronic signal (b) FFT frequency vs. probe photon energy ($h\nu$) and (c) the summation of the FFT amplitude from (b) at $B_0$ = 0 T.

## 4. Exciton Energy Modulation by Coherent Spin Waves

**CrSBr bulk crystal**. We estimate the energy modulation of excitons in the CrSBr bulk sample by adding scaled steady state reflectance to the ΔR/R spectra. Here, we subtract the incoherent electronic signal from the raw ΔR/R spectra (Fig. S2b). The resulting subtracted spectra at different probe wavelengths are shown in Fig. S5. The time-dependent spectral change of the minimum and the maximum of the subtracted ΔR/R spectra can be approximated as the magnitude of the energy modulation (Fig. S6). Taking the FFT of the energy modulation, we obtain the two magnon frequencies, 24 and 34 GHz. We quantify the magnitude of the modulated exciton energy by adding the steady state reflectance spectra to the subtracted ΔR/R spectra (Fig. S8). The phase flip in the signal can be understood as the exciton energy moving to the left or right with respect to the exciton energy at time 0. For bulk CrSBr, the measured change in exciton energy is 0.9 meV in the 1.358 eV region and 3.2 meV in the 1.340 eV region.

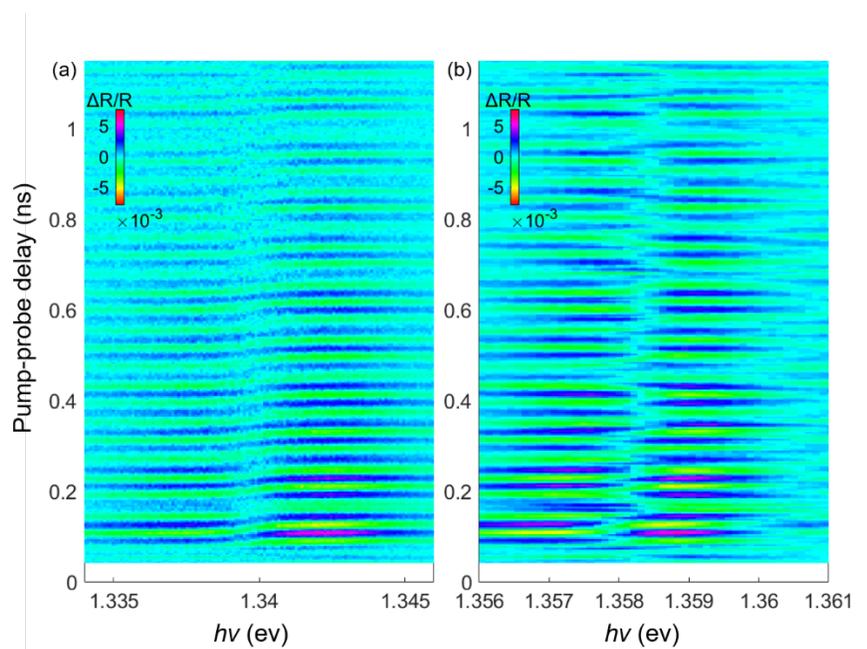

**Fig. S6.** Transient reflectance spectra around the exciton energy of (a) 1.340 eV and (b) 1.358 eV.

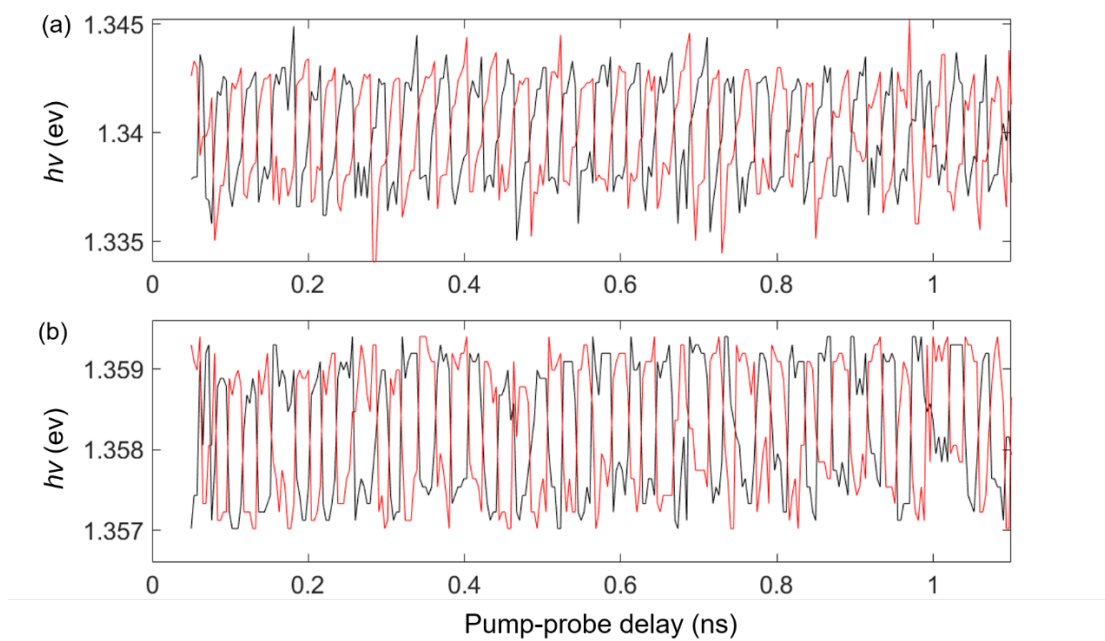

**Fig. S7.** Time dependent traces of maximum (black) and minimum (red) peak of the ΔR/R at (a) 1.358 eV and (b) 1.340 eV.

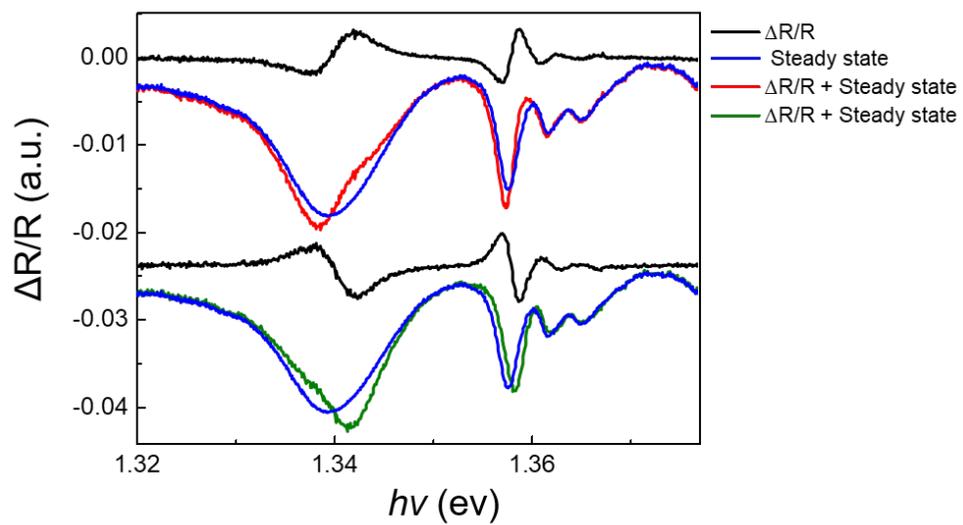

**Fig. S8.** Addition of the scaled steady state reflectance spectra to the subtracted ΔR/R spectra.

**Energy modulation in a few layer CrSBr.** The same analysis is applied to a few layers of CrSBr samples. Unlike the bulk sample, a few layers CrSBr samples contain one major exciton peak around 910 nm. Although the amplitude of magnon oscillation is large at a single probe wavelength down to the trilayer CrSBr, the amplitude is small in the bilayer. Therefore, the magnon frequencies in the bilayer sample is integrated over the entire exciton peak. Thus, the magnitude of the energy modulation cannot be identified for bilayer but magnon frequencies can be still detected.

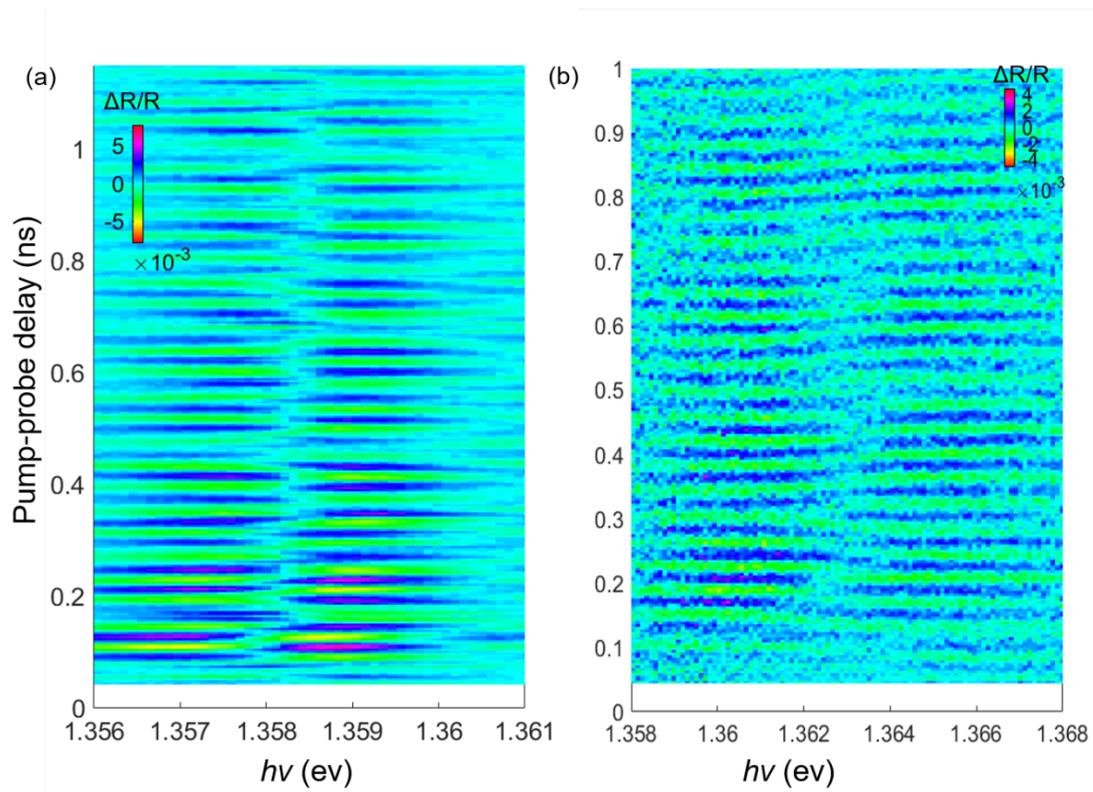

**Fig. S9.** Transient reflectance spectra of (a) 4L and (b) 3L after the subtraction of incoherent electronic signals. The corresponding spectrum for 5L is shown in the main text Fig. 4a.

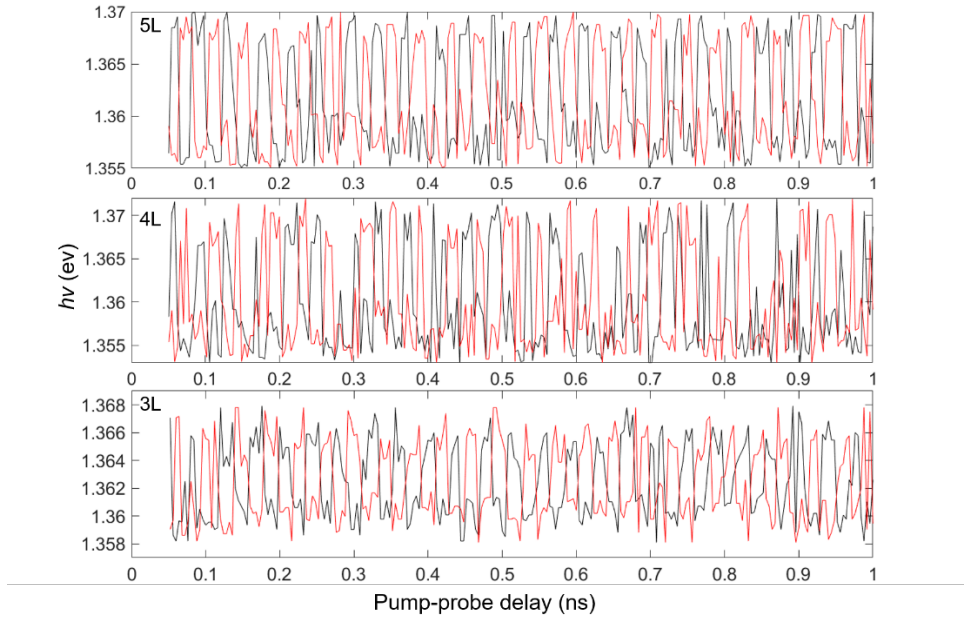

**Fig. S10.** Time dependent traces of maximum (black) and minimum (red) peak of the subtracted ΔR/R.

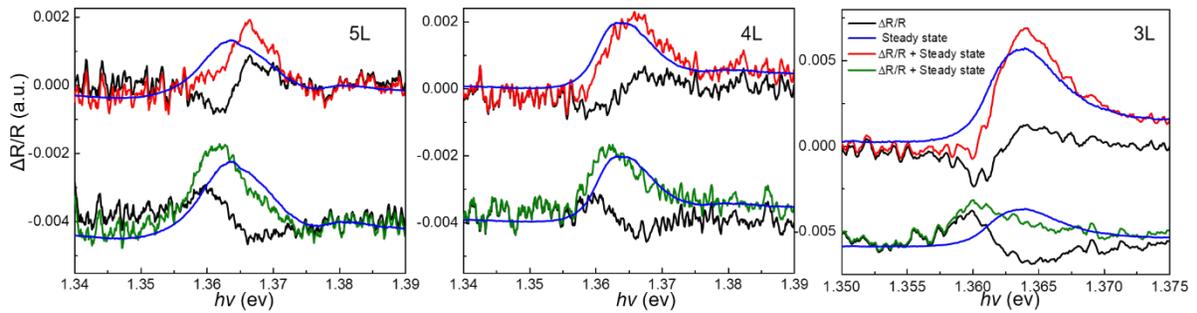

**Fig. S11.** Addition of the scaled steady state reflectance spectra to the subtracted ΔR/R spectra.

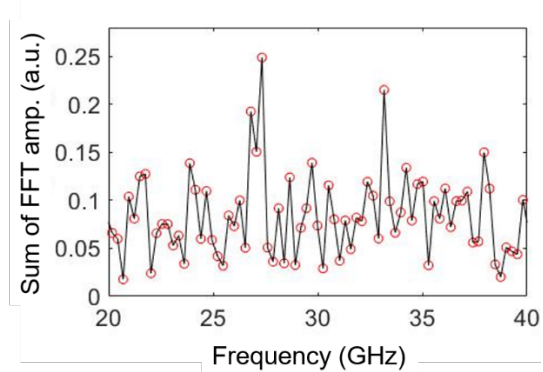

**Figs. S12.** FFT spectrum averaged over the entire probe photon energy of 2L.

Table S1. Magnon frequencies and energy modulation

| Sample | Energy modulation |
|---|---|
| Bulk | 1.340 eV (3.2 meV) 1.358 eV (0.9 meV) |
| 5L | 5.1 meV |
| 4L | 3.7 meV |
| 3L | 4.0 meV |

5. AFMR Data Collection and Analysis

**AFMR measurement.** Antiferromagnetic resonance spectra are collected by using a broadband coplanar waveguide method in a Quantum Design Physical Property Measurement system at variable temperature and a variable magnetic field. An exfoliated CrSBr bulk sample (5 mm by 2mm) is mounted on a coplanar waveguide where an ac current provides a small oscillating Oersted field that drives the magnetization into a small angle precession. At the resonance condition, the amplitude of precession is maximized. An external dc magnetic field is applied perpendicular to the sample plane. The frequency of ac field is fixed while the dc magnetic field is swept from 3T to zero with a step size of 0.02 T. Both the transmission and reflection spectra are collected using a vector network analyzer (VNA). The real part from the transmission port is plotted with respect to the swept magnetic field (Fig. S13). A linear combination of symmetric and antisymmetric Lorentzian functions is used to extract peak position of resonance peaks (Fig. S13).

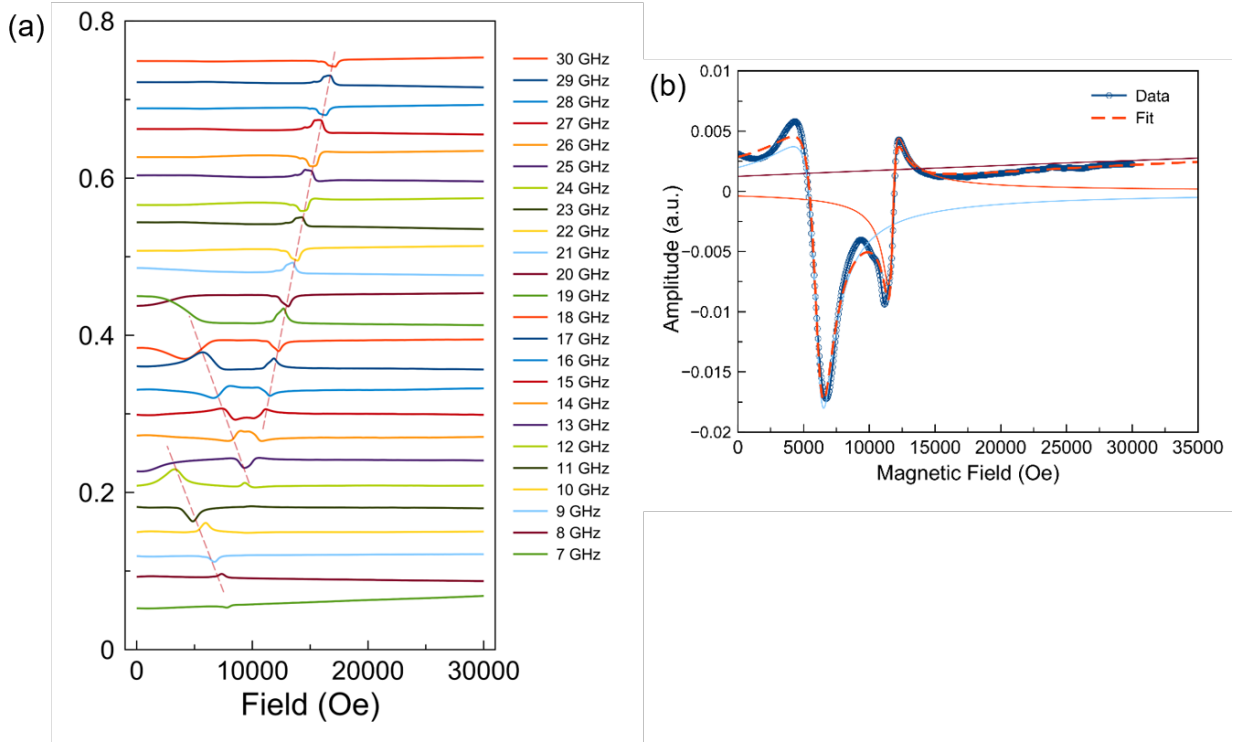

**Fig. S13.** (a) The real part of the spectra of S12, transmission port at different resonance frequencies at 100K. (b) Peak fitting using a linear combination of symmetric and antisymmetric Lorentzian functions.

6. Linear spin wave theory (LSWT)

As a semiclassical alternative approach to the classical LL equations shown below, we also calculated the magnon modes using linear spin wave theory (LSWT). Because the relevant exchange interactions for the Q=0 modes observed here do not depend upon the in-plane exchange interactions, we can model the system as a 1D antiferromagnetic spin chain with anisotropy. We write the Hamiltonian as

$$H = \sum_i J_{int} S_i \cdot S_{i+1} + A_x S_x^2 + A_z S_z^2$$

where $J_{int}$ is the net interlayer exchange interaction, and $A_x$ and $A_z$ are single ion anisotropy terms. Here we assume that $A_x$ and $A_z$ are both positive and $A_y$ is zero such that the easy axis is in the b direction. Solving this equation for the magnon modes using the Holstein-Primakoff formalism[3] gives the formulae in the main text for the two low energy modes. (Note that these modes are the acoustic intraplane magnon mode, split by the presence of interplane exchange and magnetic anisotropy. Note also that the presence of an intermediate axis is crucial to the observation of two modes at zero applied magnetic field: in a single easy-axis case $A_z = A_x$, the two B=0 modes are degenerate.

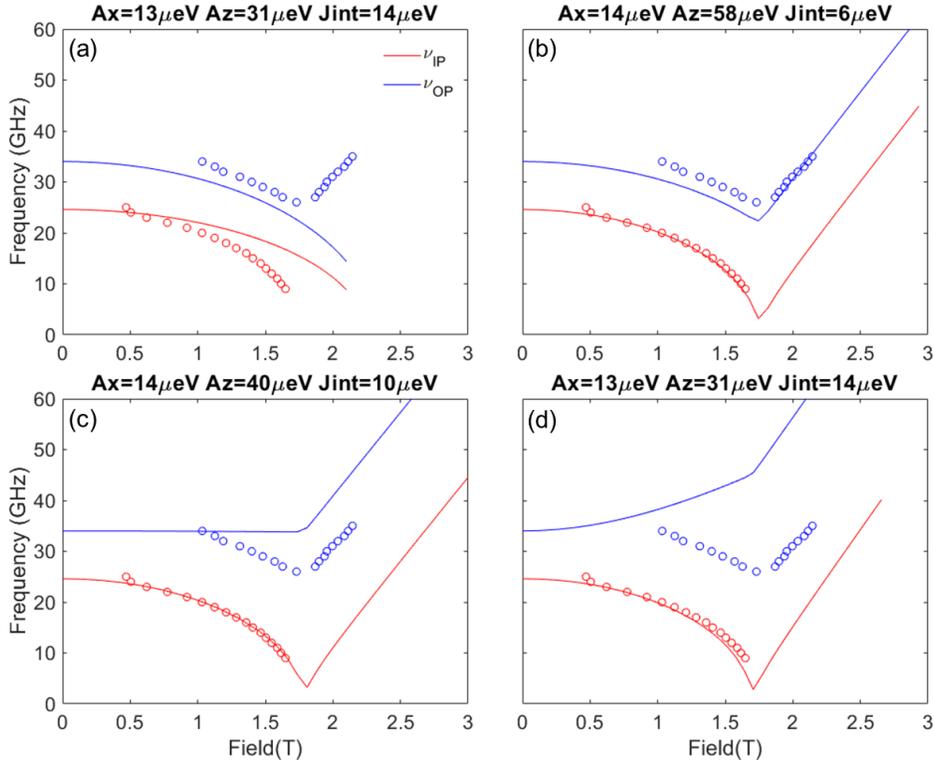

**Fig. S14.** LSWT fitting of magnetic field dependent AFMR data shown in Fig. 2b in main text using different single ion anisotropy and interlayer exchange constants.

We then model the field-dependence of the modes by minimizing the spin structure with various values of external magnetic field, and recalculating the Q=0 modes using the *SpinW* software package.[4] Because the interplane exchange is unknown, we model the field dependence using many different $J_{int}$ values, adjusting $A_z$ and $A_x$ to produce the observed zero field magnon energies (Fig. S14). We find that an interplane exchange energy below 0.04 meV is required to have both magnon modes initially decrease energy with magnetic field, with a more realistic value closer to 0.02 meV. (Note that getting the LSWT calculated transition field---the kink in magnon energy---to match experiment requires an adjusting of the *g* value to less than 2, similar to the adjustment required for the gyromagnetic ratio in the LL equations below. Thus, by having observed the energy and field dependence of the Q=0 magnon modes, we can place an upper bound on the CrSBr interplane exchange interactions: three orders of magnitude weaker than the intraplanar exchange.[5]

## 7. Analysis Based on the Landau-Lifshitz (LL) Equation

In order to understand the three branches from AFMR measurement, we modeled the dynamical spin precession using the LL equation:

$$\frac{d\vec{m}_{(i)}}{dt} = -\mu_0 \gamma \vec{m}_{(i)} \times \vec{H}_{eff_{(i)}} \quad (eq.\,1)$$

Here, $\vec{m}_{(i)}$ is the magnetic moment where $i = 1$ (top) or 2 (bottom) layer.

We first write down the total energy and use the following equation to find the $\vec{H}_{eff}$:

$$\vec{H}_{eff_{(i)}} = \frac{-1}{\mu_0 M_s} \nabla_{\vec{m}_{(i)}} E_{total} \quad (eq.\,2)$$

Since CrSBr is a triaxial, AFM system with an easy axis along the b-axis, intermediate axis along the a-axis and hard-axis along c-axis, we consider four energy contributions, Zeeman energy, two different magnetocrystalline anisotropy energy and interlayer exchange energy.

$$E_{total} = -\vec{m_1} \cdot \vec{H}_{ext,c} - \vec{m_2} \cdot \vec{H}_{ext,c} - \frac{1}{2} H_A m_{b,1}^2 - \frac{1}{2} H_A m_{b,2}^2 - \frac{1}{2} H_B m_{a,1}^2 - \frac{1}{2} H_B m_{a,2}^2$$

$$+ H_E \vec{m_1} \cdot \vec{m_2} \quad (eq.\,3)$$

Here, $H_{ext}$ is applied along the c-axis, $\vec{m_1}$ and $\vec{m_2}$ are magnetic moments in the layer 1 and layer 2, respectively, $H_A$ describes the magnetic anisotropy along the easy axis, $H_B$ along the intermediate axis and $H_E$ is the AFM interlayer exchange field (Fig. S14). $H_A$ is greater than $H_B$ because b-axis is the easy axis. We ignored a universal factor of $\mu_0 M_s$.

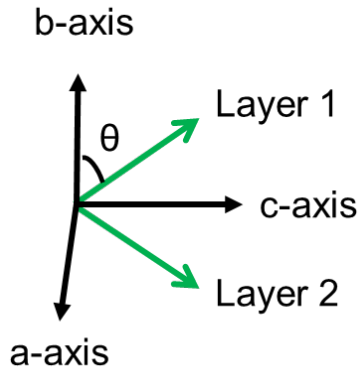

**Fig. S15.** Magnetic moments represented in crystallographic axis

Following the coordinate system shown in Figure S12, the total energy can be written down as:

$$E_{total} = -2H_{ext} \sin\theta - H_A \cos^2\theta - H_E \cos 2\theta - H_B (\sin(\theta)\sin(\varphi))^2 \quad (eq.4)$$

Taking the derivative with respect to $\theta$ and $\varphi$,

$$\frac{dE_{total}}{d\theta} = -2H_{ext}\cos(\theta) + 2H_A \cos(\theta)\sin(\theta) \, 4H_E \cos(\theta)\sin(\theta)$$
$$- 2H_B \sin(\theta)\cos(\theta)\sin^2(\varphi) \quad (eq.5)$$

$$\frac{dE}{d\varphi} = 2H_B \sin^2(\theta)\sin(\varphi)\cos(\varphi) \quad (eq.6)$$

At equilibrium, $\frac{dE_{total}}{d\theta} = 0$ and $\frac{dE_{total}}{d\varphi} = 0$ gives

$$\sin\theta = \frac{H_{ext}}{2H_E + H_A} \quad (eq.7)$$

$$\frac{dE}{d\varphi} = 2H_B \sin^2(\theta)\sin(\varphi)\cos(\varphi) = 0 \quad (eq.8)$$

Thus $\varphi = 0$ at equilibrium.

Using the eq. 2, $\overrightarrow{H_{eff}}$ for each layer can be written as the following where i and j refer to top and bottom layer:

$$\overrightarrow{H_{eff(\iota=1,2)}} = \overrightarrow{H_{ext,c}} - H_E \overrightarrow{m_j} + H_A m_{b,i} \hat{b} + H_B m_{a,i} \hat{a} \quad (eq.9)$$

We will first find the eigenfrequencies when the out-of-plane field is large enough to saturate both magnetization, $H_{ext,c} > H_s$. In this case, the magnetic moment in AFM coupled layers can be written as, $m_1 \approx (m_{a,1} \; 1 \; m_{b,1})$ and $m_2 \approx (m_{a,2} \; 1 \; m_{b,2})$.

Then the effective field and the time derivative of magnetic moment can be written as,

$$\overrightarrow{H_{eff(1)}} = H_{ext,c}\begin{pmatrix}0\\1\\0\end{pmatrix} - H_E \begin{pmatrix}m_{a,2}\\1\\m_{b,2}\end{pmatrix} + H_A \begin{pmatrix}0\\0\\m_{b,1}\end{pmatrix} + H_B \begin{pmatrix}m_{a,1}\\0\\0\end{pmatrix} \quad (eq.9)$$

$$\frac{d\vec{m_1}}{dt} = \begin{pmatrix} m_{a,1} \\ 0 \\ m_{b,1} \end{pmatrix} \quad (eq.10)$$

where the magnetization has the form of $e^{-i\omega t}$. Plugging these into the eq.1 without a damping term and ignoring the 2nd order and higher terms we get the 4 x 4 matrix:

$$\frac{i\omega}{\mu_0 \gamma} \begin{pmatrix} m_{a,1} \\ m_{b,1} \\ m_{a,2} \\ m_{b,2} \end{pmatrix} = \begin{pmatrix} 0 & H_A - H_{ext} + H_E & 0 & -H_E \\ H_{ext} - H_E - H_B & 0 & H_E & 0 \\ 0 & -H_E & 0 & H_A - H_{ext} + H_E \\ H_E & 0 & H_{ext} - H_E - H_B & 0 \end{pmatrix} \quad (eq.11)$$

Moving the left term to the right and taking the determinant equals to 0, we can get two eigenfrequencies:

$$f_1 = \frac{\mu_0 \gamma}{2\pi} \sqrt{(H_A + 2H_E - H_{ext})(H_B + 2H_E - H_{ext})} \quad (eq.12)$$

$$f_2 = \frac{\mu_0 \gamma}{2\pi} \sqrt{(H_A - H_{ext})(H_B - H_{ext})} \quad (eq.13)$$

where $f_1$ and $f_2$ corresponds to the in-phase and out-of-phase precession above $H_s$.

When the external field is below the saturation field, the magnetic moment $m_1$ and $m_2$ have the opposite sign. Assuming that $m_1>0$ and $m_2<0$, and the sub-lattice 1 is on the top and sub lattice 2 is at the bottom, we can write magnetic moment in the each other's frame by applying rotation matrix of sublattice 2 respective to sublattice 1:

$$\Gamma_{2,1} = \begin{pmatrix} 1 & 0 & 0 \\ 0 & -\cos(2\theta) & +\sin(2\theta) \\ 0 & -\sin(2\theta) & -\cos(2\theta) \end{pmatrix} \quad (eq.14)$$

Similarly, the rotation matrix from sublattice 1 to sublattice 2 is:

$$\Gamma_{2,1} = \begin{pmatrix} 1 & 0 & 0 \\ 0 & -\cos(2\theta) & +\sin(2\theta) \\ 0 & +\sin(2\theta) & -\cos(2\theta) \end{pmatrix} \quad (eq.\,15)$$

Applying the rotation matrix to the magnetic moment and solving the eq.1 without a damping term, we obtain the 4 x 4 matrix:

$$\frac{i\omega}{\mu_0 \gamma} \begin{pmatrix} m_{a,1} \\ m_{b,1} \\ m_{a,2} \\ m_{b,2} \end{pmatrix} = \begin{pmatrix} 0 & -H_E - H_A \cos^2(\theta) & 0 & H_E \cos(2\theta) \\ H_E + H_A - H_B & 0 & H_E & 0 \\ 0 & H_E \cos(2\theta) & 0 & -H_E - H_A \cos^2(\theta) \\ H_E & 0 & H_E + H_A - H_B & 0 \end{pmatrix} \quad (eq.\,16)$$

Once again, moving the left term to the right and taking the determinant equals to 0 and using the eq 7, we get the two eigenfrequencies:

$$f_1 = \frac{\mu_0 \gamma}{2\pi} \sqrt{(H_A - H_B)(H_A + 2H_E)\left(1 - \left(\frac{H_{ext}}{(H_A + 2H_E)}\right)^2\right)} \quad (eq.\,17)$$

$$f_2 = \frac{\mu_0 \gamma}{2\pi} \sqrt{(H_A - H_B + 2H_E)\left(2H_E - H_A\left(\left(\frac{H_{ext}}{(H_A + 2H_E)}\right)^2 - 1\right)\right) + 2H_E\left(\left(\frac{H_{ext}}{(H_A + 2H_E)}\right)^2 - 1\right)} \quad (eq.\,18)$$

where $f_1$ and $f_2$ corresponds to the in-phase and out-of-phase precession below $H_s$.

At 0 magnetic field, the two eigenfrequencies obtained from LL, (eq. 17) and (eq. 18), become $f_1 = \frac{\mu_0 \gamma}{2\pi} \sqrt{(H_A - H_B)(H_A + 2H_E)}$ and $f_2 = \frac{\mu_0 \gamma}{2\pi} \sqrt{H_A(H_A - H_B + 2H_E)}$ for $v_{IP}$ and $v_{OP}$, respectively. These equations are complementary to the two eigenfrequencies obtained from LSWT, $2S\sqrt{A_x A_z + A_x J_{int}}$ and $2S\sqrt{A_x A_z + A_z J_{int}}$, since $H_A$, $H_B$, and $J_{int}$ are proportional to ($A_z$- $A_y$), ($A_z$-$A_x$) and $2H_E$, respectively. Since $A_y$ is set to 0 during LSWT fitting, we can recover the same equations by replacing anisotropy and exchange field to single ion anisotropy and exchange constants.

Although we expect to see 4 branches as described in the 4 eigenfrequencies, we only observe three branches without the in-phase precession branch after saturation field. This is persistent throughout all temperature ranges. We suspect that due to some symmetry constraint, we cannot detect this branch, similar to that observed in CrCl$_3$.[6] We fit three branches using the three eigenfrequencies described in eq 12, 17 and 18. We obtain consistent exchange ($H_E$) and anisotropy fields ($H_A$ and $H_B$) from fitting different branches and they match well with the measured anisotropy fields from magnetic isotherm (Table S4). The electron gyromagnetic ratio term, $\frac{\mu_0 \gamma}{2\pi}$ slightly deviates from the expected value, 28 GHz/T at different temperatures (Tables S2-S4). Presumably, ignoring higher order terms could lead to slightly lower value of gyromagnetic ratio. This is also observed in LSWT fitting as discussed above.

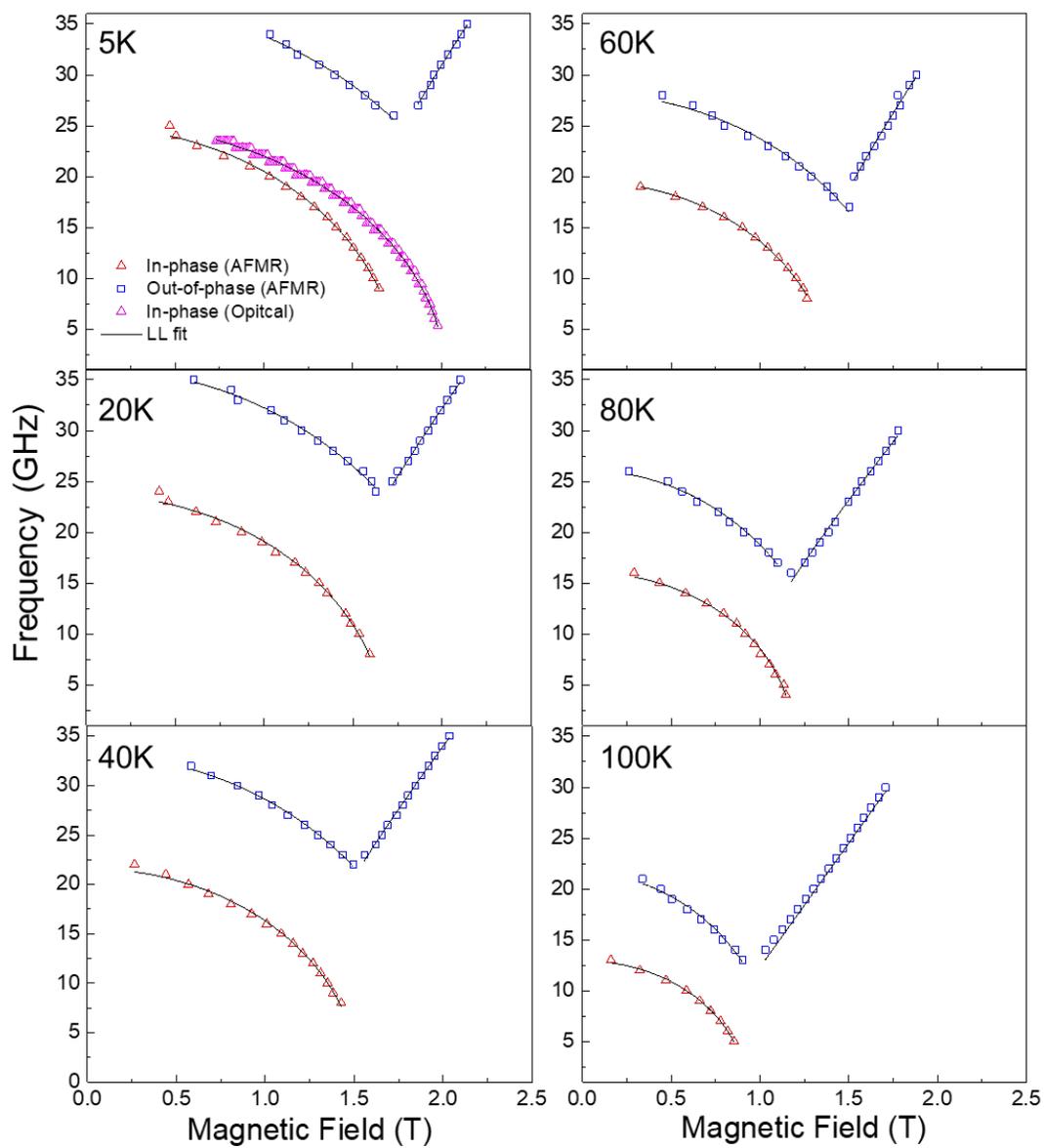

**Fig. S16.** AFMR frequency vs. magnetic field plots at different temperature. Note, the saturation field is decreasing with the temperature.

Table S2. LLG fit parameters for the out-of-phase precession below the saturation field

| Temperature (K) | $\gamma/2\pi$ (T) | $H_E$ (T) | $H_A$ (T) | $H_B$ (T) |
|---|---|---|---|---|
| 5 | 24.9 | 0.26 | 1.60 | 0.64 |
| 20 | 28 | 0.22 | 1.3 | 0.5 |
| 40 | 27 | 0.14 | 1.5 | 0.8 |
| 60 | 26 | 0.13 | 1.4 | 0.8 |
| 80 | 27 | 0.12 | 1.0 | 0.3 |
| 100 | 28 | 0.11 | 0.69 | 0.2 |

Table S3. LLG fit parameters for the in-phase precession below the saturation field

| Temperature (K) | $\gamma/2\pi$ (T) | $H_E$ (T) | $H_A$ (T) | $H_B$ (T) |
|---|---|---|---|---|
| 5 (AFMR) | 16.8 | 0.27 | 1.6 | 0.50 |
| 5 (Optical) | 28 | 0.16 | 1.4 | 1.0 |
| 20 | 28 | 0.24 | 1.2 | 0.77 |
| 40 | 25 | 0.20 | 1.1 | 0.64 |
| 60 | 21 | 0.10 | 1.2 | 0.6 |
| 80 | 27 | 0.2 | 0.78 | 0.48 |
| 100 | 21 | 0.10 | 0.72 | 0.33 |

Table S4. LLG fit parameters for the out-of-phase precession above the saturation field

| Temperature (K) | $\gamma/2\pi$ (T) | $H_A$ (T) | $H_B$ (T) |
|---|---|---|---|
| 5 | 23 | 1.3 | 0.5 |
| 20 | 21 | 1.2 | 0.8 |
| 40 | 22 | 1.0 | 0.46 |
| 60 | 22 | 1.2 | 0.65 |
| 80 | 21 | 0.83 | 0.34 |
| 100 | 24 | 0.53 | 0.44 |

8. Vibrating Sample Magnetometry

Vibrating sample magnetometry (VSM) measurements were conducted on a Quantum Design PPMS DynaCool system. Two single crystals of CrSBr were co-aligned and attached to a quartz paddle using GE varnish with the crystallographic a-axis aligned parallel to the length of the quartz paddle, which coincides with the direction of the applied field. Variable-field magnetization curves were collected from 0 to 5 T (20 Oe/s) at the temperatures listed in Table S5. For each temperature, one linear fit was applied to the high-field data ($H > 2$ T), above the saturation field, and a second linear fit was applied to the linear region at low fields. The intersection of these two lines yields the anisotropy field for the *a*-axis as its abscissa. The crystals were then removed from the quartz paddle using a 1:1 ethanol/toluene solution and re-attached to the paddle with the *c*-axis aligned parallel to the field direction. The same measurement and data analysis procedures were then applied to obtain the c-axis anisotropy fields.

Table S5. Anisotropy field from temperature dependent magnetization measurement with the magnetic field applied along *a*- and *c*- axis.

| Temperature (K) | $H_A$-$H_B$ (*a*-axis) T | $H_A$ (*c*-axis) T |
|---|---|---|
| 2 | 1.054 | 1.661 |
| 10 | 1.027 | 1.647 |
| 20 | 0.979 | 1.597 |
| 30 | 0.924 | 1.532 |
| 40 | 0.887 | 1.474 |
| 50 | 0.827 | 1.397 |
| 60 | 0.771 | 1.330 |
| 70 | 0.723 | 1.243 |
| 80 | 0.655 | 1.156 |
| 90 | 0.592 | 1.071 |
| 100 | 0.524 | 0.965 |
| 110 | 0.480 | 0.852 |
| 120 | 0.388 | 0.731 |
| 130 | 0.284 | 0.582 |

9. Coherent Spin Wave Propagation

The probe-wavelength dependent short time Fourier transform (STFT) is performed at a fixed frequency of 2 GHz, 24 GHz and 34 GHz along both a and b-axis. Using the MATLAB built in STFT function, time window of 600 ps and a hamming window of 590 ps are used for 2 GHz mode. For 24 GHz and 34 GHz, time window of 150 ps with a hamming window of 140 ps is used. The rise and decay of the STFT indicates that magnons are propagating coherently. At a longer probe distance, the width of the amplitude gets broader and the amplitude decays as the propagating magnons lose temporal and spatial coherence. The group velocities of magnon and phonon are obtained from the linear fitting of the pump-probe distance (d) v.s. pump-probe time delay (Fig. S17).

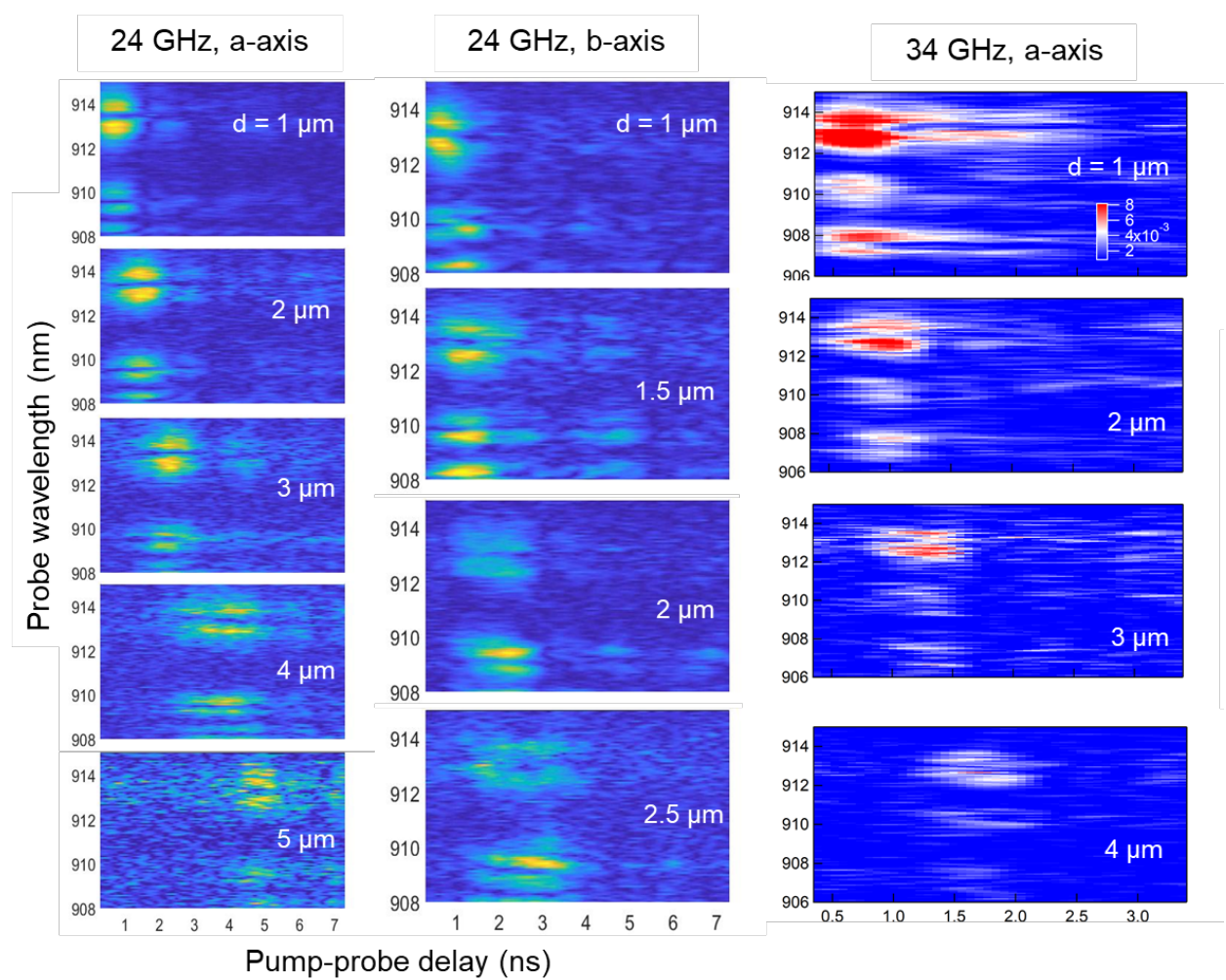

**Fig. S17.** STFT spectra of 24 GHz and 34 GHz magnon modes along the *a*- and *b*-axis of the bulk CrSBr.

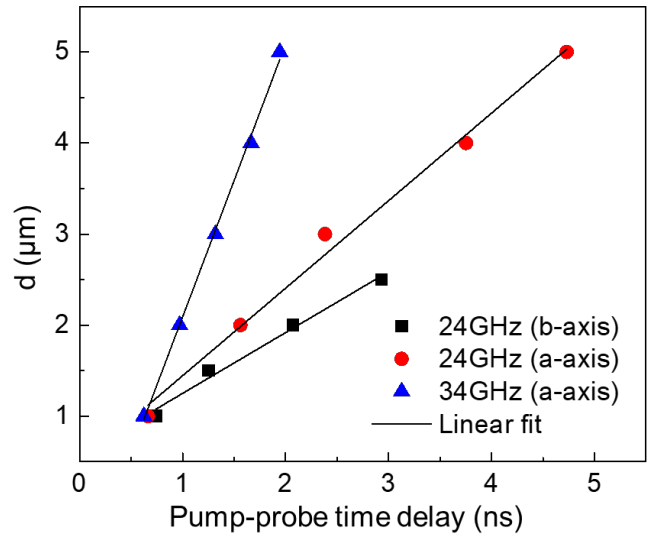

**Fig. S18.** Pump-probe distance (d) vs. time delay for magnon modes with group velocities of x, y, and z for 24 GHz mode along a and b axis and for 34 GHz mode along the a-axis.

10. FFT Peak Width Analysis

We determine the coherence lifetime from the full-width-half-maximum (FWHM) of the Lorentzian peak fitting. The time step is 4 ps and the number of data points are 1950, thus the frequency resolution is 0.1 GHz. From Lorentzian peak fitting, we obtain three peaks, 25.7 ± 0.1, 24.5 ± 0.2, and 34.7 ± 0.2. The FWHM of the three peaks corresponds to 0.2 GHz, 0.4 GHz and 0.4 GHz which translate into 5 ns, 2.5 ns and 2.5 ns.

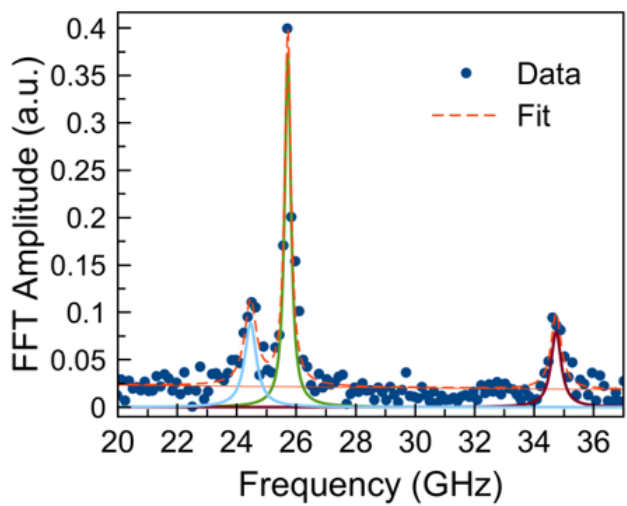

**Fig. S19.** Lorentzian peak fitting of the FFT spectra averaged over the entire probe photon energy.

## 11. Magnon Peak Splitting

As discussed in the main text, magnon peaks split into two peaks and the width of the two peaks depends on the probe wavelength. For instance, at 908.5 nm 910.4 nm and 912.9 nm, the splitting width is 0.93 GHz, 0.93 GHz and 0.4 GHz for 24 GHz and 0.4 GHz, 1 GHz and 0.53 GHz for 34 GHz.

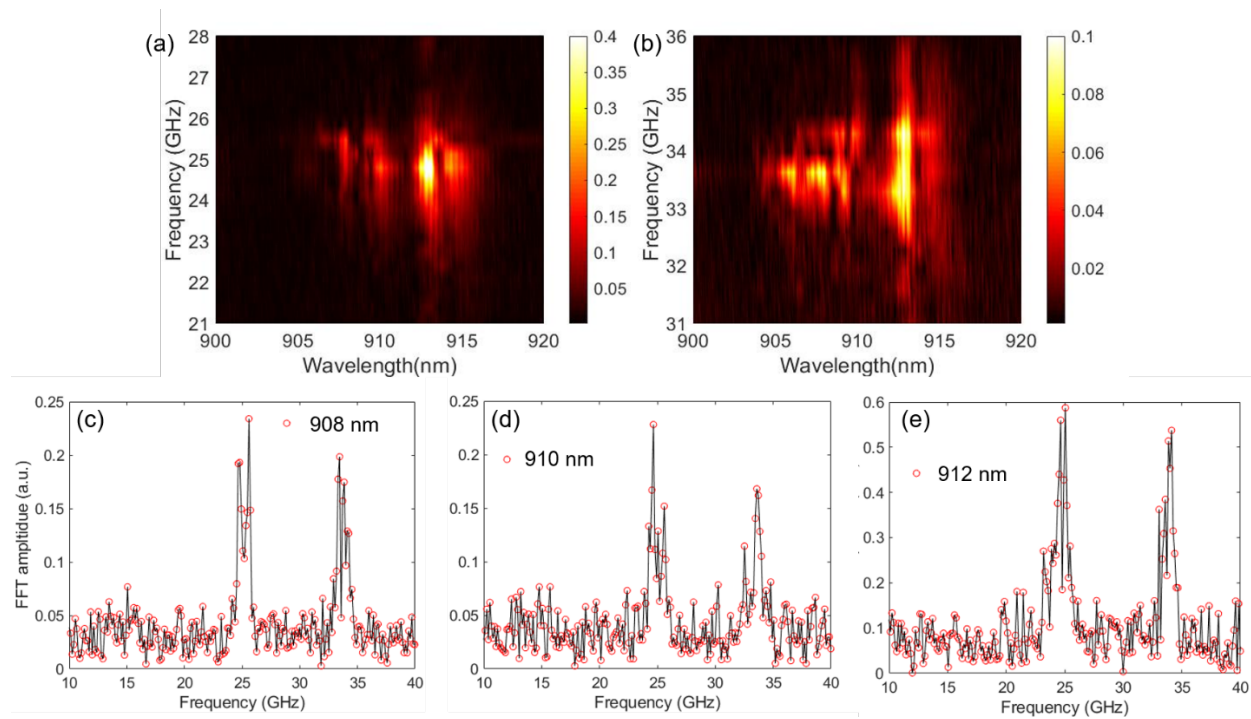

**Fig. S20.** FFT spectra of the two magnon. (a) 24 GHz and (b) 34 GHz modes show peak splitting at (c), (d), and (e) different probe wavelengths.

## 12. Coherence time for 5L sample

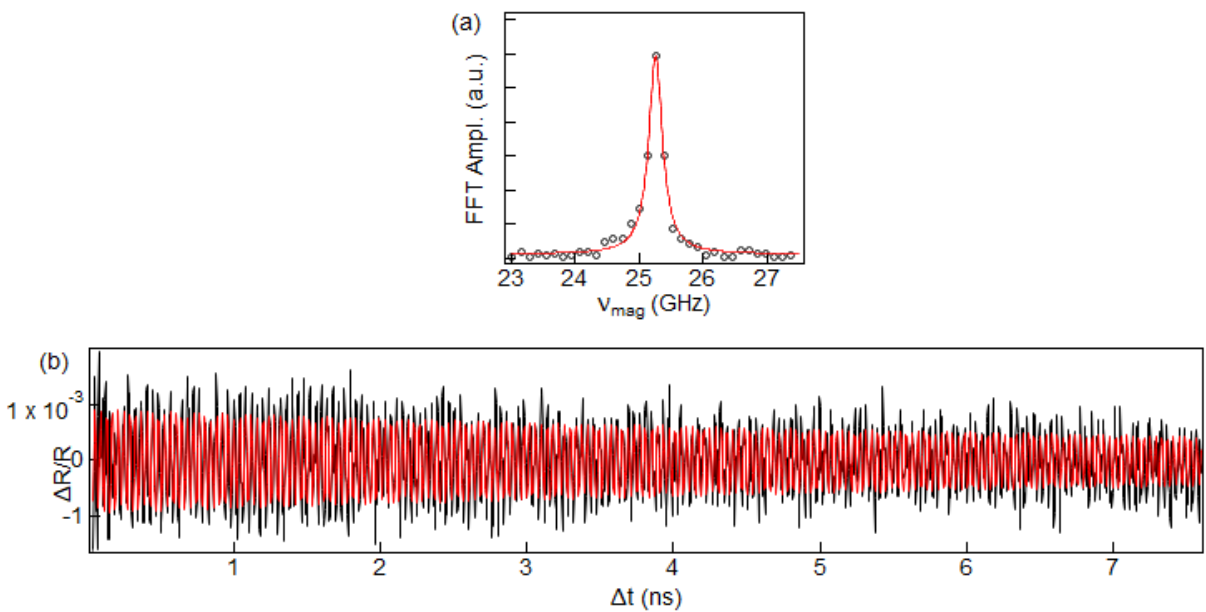

**Fig. S21.** FFT spectrum of the (a) 24 GHz mode of the 5L sample discussed in main text Fig. 5b and (b) time domain oscillating signal (black) with damped sine wave fitting (red). FFT peak fitting gives a full-width-at-half-maximum of 0.20±0.02 GHz, which corresponds to decoherence time of 5 ±0.5 ns and damped sine wave fitting gives decoherence time of 10 ± 1 ns.